\documentclass{article} % For LaTeX2e
\usepackage{iclr2025_conference,times}
\usepackage{booktabs}
\usepackage{amsmath,amsfonts,bm}

\def\eqref#1{equation~\ref{#1}}
\def\1{\bm{1}}

\DeclareMathAlphabet{\mathsfit}{\encodingdefault}{\sfdefault}{m}{sl}
\SetMathAlphabet{\mathsfit}{bold}{\encodingdefault}{\sfdefault}{bx}{n}

\usepackage{hyperref}
\usepackage{url}
\usepackage{graphicx}
\usepackage{svg}
\usepackage{float}
\usepackage{booktabs}
\usepackage{caption}
\usepackage{listings}

\title{From Intuition to Understanding: Using AI Peers to Overcome Physics Misconceptions}

\author{\hspace{-0.8mm}
\textbf{Ruben Weijers\textsuperscript{1}}
 \And
 \textbf{Denton Wu\textsuperscript{2}}
 \And
 \textbf{Hannah Betts\textsuperscript{3}}
 \And
  \textbf{Tamara Jacod\textsuperscript{3}}
 \And
  \textbf{Luke Yuxiang Guan\textsuperscript{3}}
 \And
  \textbf{Vidya Sujaya\textsuperscript{4,5}}
 \And
 \textbf{Kushal Dev\textsuperscript{3}} 
 \And
 \textbf{Toshali Goel\textsuperscript{3}} 
 \And
 \textbf{William Delooze\textsuperscript{2}}
 \And
 \textbf{Reihaneh Rabbany\textsuperscript{4,5}}
 \And
 \textbf{Ying Wu\textsuperscript{2}}
 \And
 \textbf{Jean-François Godbout\textsuperscript{4,6}}
 \And
 \textbf{Kellin Pelrine\textsuperscript{4,5}}
\AND
\normalfont
 \textsuperscript{1}Utrech University\hspace{0.8mm}
 \textsuperscript{2}Duke University\hspace{0.8mm}
 \textsuperscript{3}Independent\hspace{0.8mm}\\
 \textsuperscript{4}Mila Institute\hspace{0.8mm}
 \textsuperscript{5}McGill University
 \textsuperscript{6}Université de Montréal\hspace{0.8mm}
\\
 \small{
   \textbf{Correspondence:} \href{kellin.pelrine@mila.quebec}{kellin.pelrine@mila.quebec}
 }
}

\iclrfinalcopy % Uncomment for camera-ready version, but NOT for submission.
\begin{document}

\maketitle

%%%%%%%%%%%%%%%%%%%%%%%%%%%%%%%%%%%%%%%%%%%%%%%%%%%%%%%
%%%%%%%%%%%%%% ABSTRACT SECTION %%%%%%%%%%%%%%%%%%%%%%%
%%%%%%%%%%%%%%%%%%%%%%%%%%%%%%%%%%%%%%%%%%%%%%%%%%%%%%%
\begin{abstract}

Generative AI has the potential to transform personalization and accessibility of education. However, it raises serious concerns about accuracy and helping students become independent critical thinkers. In this study, we designed a helpful AI ``Peer'' to help students correct fundamental physics misconceptions related to Newtonian mechanic concepts. In contrast to approaches that seek near-perfect accuracy to create an authoritative AI tutor or teacher, we directly inform students that this AI can answer up to 40\% of questions incorrectly. In a randomized controlled trial with 165 students, those who engaged in targeted dialogue with the AI Peer achieved post-test scores that were, on average, 10.5 percentage points higher—with over 20 percentage points higher normalized gain—than a control group that discussed physics history.  Qualitative feedback indicated that 91\% of the treatment group's AI interactions were rated as helpful. Furthermore, by comparing student performance on pre- and post-test questions about the same concept, along with experts' annotations of the AI interactions, we find initial evidence suggesting the improvement in performance does not depend on the correctness of the AI. With further research, the AI Peer paradigm described here could open new possibilities for how we learn, adapt to, and grow with AI.

\end{abstract}

%%%%%%%%%%%%%%%%%%%%%%%%%%%%%%%%%%%%%%%%%%%%%%%%%%%%%%%%%
%%%%%%%%%%%%%% INTRODUCTION SECTION %%%%%%%%%%%%%%%%%%%%%
%%%%%%%%%%%%%%%%%%%%%%%%%%%%%%%%%%%%%%%%%%%%%%%%%%%%%%%%%
\section{Introduction}

Students have recently been exposed to the remarkable capabilities of Generative AI (AI) in education (AIED). For example, OpenAI's ChatGPT has been reported to successfully support teaching preparation, assessment design and grading, and student learning \citep{lo2023impact}. Systems like ChatGPT show potential to save time and enhance teaching and learning, including critical and higher-order thinking tasks \citep{lo2023impact}.

However, there is limited concrete evidence whether these tools are effective at improving student learning outcomes \citep{samson2025curve}. In fact, LLMs are well-reported to hallucinate information and provide sycophantic answers \citep{wei2023simple, perez-etal-2023-discovering}, suggesting that AI could actually be detrimental to student learning if it introduces inaccuracies and biases into classroom materials. Given that an increasingly large number of students are now using AI to help them with their school assignments \citep{stryker2024poll}, there is a growing concern that students must be taught `AI literacy' to recognize and evaluate potential errors generated by LLMs \citep{Wineburg2024student-misinfo}. Although recent techniques such as Retrieval-Augmented Generation \citep{piktus2020rag} have improved the veracity of generated content, experts disagree on whether hallucinations will be consistently preventable even in the future \citep{samson2025curve}---as they are clearly not today. Thus, there is an urgent need to determine whether AIED can be an effective tool for education despite these inherent limitations.

This research evaluates the potential of LLMs to support student learning through text-based conversations in an introductory university physics class. The study explores how LLMs can be used to facilitate learning of physics concepts, positioning the AIED as a non-expert peer rather than an expert teacher. Students complete a modified Force Concept Inventory (FCI) as a pre-test. Our AI Peer is prompted with the mistakes made by each student, resulting in a personalized discussion focused on the identified misconceptions. Students then completed the standard FCI as a post-test that asesses the same concepts via a different questionnaire. Our results demonstrate that the treatment group, after a focused discussion with the AI, experienced significantly higher learning gain between tests compared to the control group, who discussed irrelevant content (a quiz on historical figures in physics) with the AI Peer. Importantly, our AI, which leverages the full abilities of GPT-4o without any artificial reduction, was not a reliable source of truth; when tested, it answered up to 40\% of the FCI questions incorrectly. Thus, this study marks a first step towards achieving a comprehensive individualized approach to AI-assisted education by aligning human expectations to current limitations of generative AI.

Our key contributions include:

\begin{itemize}
    \item An experimental framework for measuring AI's educational capabilities. In our experiment, we create a new FCI\footnote{\href{https://github.com/ComplexData-MILA/AI4Ed}{https://github.com/ComplexData-MILA/AI4Ed}} and focus on physics. The framework can be extended to other domains and other AI systems. 
    \item Showing that AI can help correct common, deep-seated physics misconceptions that persisted over the first half of a semester of traditional instruction.
    \item A ``peer'' rather than ``instructor'' approach that we find preserves learning, while lowering barriers in how much AI accuracy is needed. With further research, this could potentially expand the areas in which we can benefit from AI, while empowering the development of human critical thinking skills.
\end{itemize}

%%%%%%%%%%%%%%%%%%%%%%%%%%%%%%%%%%%%%%%%%%%%%%%%%%%%%%%
%%%%%%%%%%%%%% LITERATURE REVIEW SECTION %%%%%%%%%%%%%%
%%%%%%%%%%%%%%%%%%%%%%%%%%%%%%%%%%%%%%%%%%%%%%%%%%%%%%%
\section{Literature Review}
\label{sec:literature}

Educators, policymakers, EdTech companies, and students are generally optimistic about the potential for AI to deliver personalized education, despite the limited evidence that these tools can be effective at improving student learning outcomes \citep{samson2025curve}. In a systematic review of 113 papers relating to AI Education tools, \citet{Chiu2023Systematic} found that, while AIED can generally improve student motivation and output, methods used to evaluate AIED were often ineffective at measuring students' learning. The authors suggested that further research is needed to ``devise new methods for evaluating the success of AI systems.'' 
They also noted that, despite the potential of AIED to provide equitable education through personalized feedback, AI could potentialy worsen educational inequity if the AIED design process lacked consideration of pedagogy and learning sciences.

Pedagogical research has identified numerous teaching strategies with the potential to significantly accelerate student learning, as described by \citet{hattie2009visible} in a synthesis of meta-analyses on student achievement. Among them are discussion with another student (argumentation), and prompt oriented or directional feedback. \citet{baidoo2023education} confirm that AI tools can enhance teaching and learning experiences by supporting personalized and interactive learning. For instance, students could leverage the capabilities of advanced generative AI to systematically explain complex concepts. \citet{mollick2023assigning} suggest several teaching strategies that could potentially enhance student learning in the presence of AI while mitigating the associated risks of these technologies. Their research emphasizes the importance of maintaining human involvement in the educational process and positioning AI as a supportive tool rather than a substitute for human instructors. 
As \citet{de2023let} argue, an AIED that combines direct instruction with inquiry should be more effective at explaining new concepts than having the AI directly answer student questions. This is precisely the approach taken by \citet{jurenka2024LearnLM}, who tested an LLM-based support tool with a class of 113 Arizona State University students. The tool, HallMate, could discuss course videos, direct learners to relevant content, provide scaffolded homework help, and assist with time management and broad learning strategies. The study tracked whether students used HallMate, how they used it, and whether students felt it was useful. It also evaluated the pedagogical quality of the conversations. While promising, this study focused on usability and engagement but did not attempt to measure the impact of AI tools on students' learning or performance.

To summarize, existing studies on the role of AI in education tend to emphasize its potential for improving teaching and learning experiences. While these studies aknowledge that AI is potentially revolutionary in education---specifically because it can provide personalized content focused on each student's learning needs---they lack robust evaluation methods for assessing AIED's impact on student learning outcomes.

A common tool found in the education literature to evaluate students' learning in a specified discipline is the `Concept Inventory' \citep{sands2018measure}. This approach consists of a set of multi-choice questions focused on testing conceptual understanding. In these evaluations, incorrect answers---called distractors---align with common student misconceptions. Thus, because they are standardized, these types of tests can measure the effectiveness of an instructional method by comparing student scores on matched questions before and after instruction.

The first concept inventory test---the `Force Concept Inventory' (FCI)---aimed to elicit students' misconceptions about fundamental Newtonian mechanics that were generally deep-held and therefore difficult to correct through conventional physics instruction \citep{halloun1985initial, hestenes1992force}. Given \citet{costello2024durably}'s recent success in reducing deep-held beliefs in conspiracy theories through dialogue with AI, we theorize that this application will transfer into education; that is, discussion with generative AI may reduce students' deep-held erroneous beliefs about fundamental Newtonian mechanics. We draw inspiration from that work for some aspects of the experimental setup, such as having students explain their reasoning in their own words. Meanwhile, the FCI is a robust tool that lets us both identify misconceptions and evaluate the effectiveness of AI for this application.  

One challenge with this approach to measuring the effectiveness of general purpose AI as a teaching tool is its propensity to produce false explanations and to hallucinate. Similarly, there is growing concern that students lack the `digital literacy' required to critically evaluate online content, including outputs of generative AI \citep{Wineburg2024student-misinfo}. One key element of digital literacy is determining the credibility of the source \citep{mcgrew2023digital-literacy}. \citet{krupp2023unreflectedacceptanceinvestigating}'s study in AI-assisted physics education demonstrated how an overreliance on generative AI led to worse student performance, even among students with a significant background in the field. Students in the study also struggled to accurately assess the correctness of LLM answers — highlighting, as the researchers noted, a need to investigate interactions that increase awareness of the imperfect reliability of LLMs.

Our primary goal in this study is to evaluate the effectiveness of AI as a tool for addressing misconceptions in an educational context. Our approach is further aimed to work even with a significant chance of inaccurate AI responses, and uses the FCI as a robust and trusted scientific measure of learning outcomes. We theorize that generative AI will show potential as a tool to effectively address student misconceptions in physics education, despite its well known limits. The results of this study can support future AIED tools to be evaluated for effectiveness at improving student learning, as well as contribute to the growing body of literature exploring the use of general purpose AI to fight other types of misconceptions, such as disinformation and conspiratorial beliefs \citep{demartini2020intheloop}. 

%%%%%%%%%%%%%%%%%%%%%%%%%%%%%%%%%%%%%%%%%%%%%%%%%
%%%%%%%%%%%%%% METHODOLOGY SECTION %%%%%%%%%%%%%%
%%%%%%%%%%%%%%%%%%%%%%%%%%%%%%%%%%%%%%%%%%%%%%%%%
\section{Methodology}
\label{sec:methodology}

\begin{figure}[h]
\begin{center}
%\framebox[4.0in]{$\;$}
\includegraphics[width=1\linewidth]{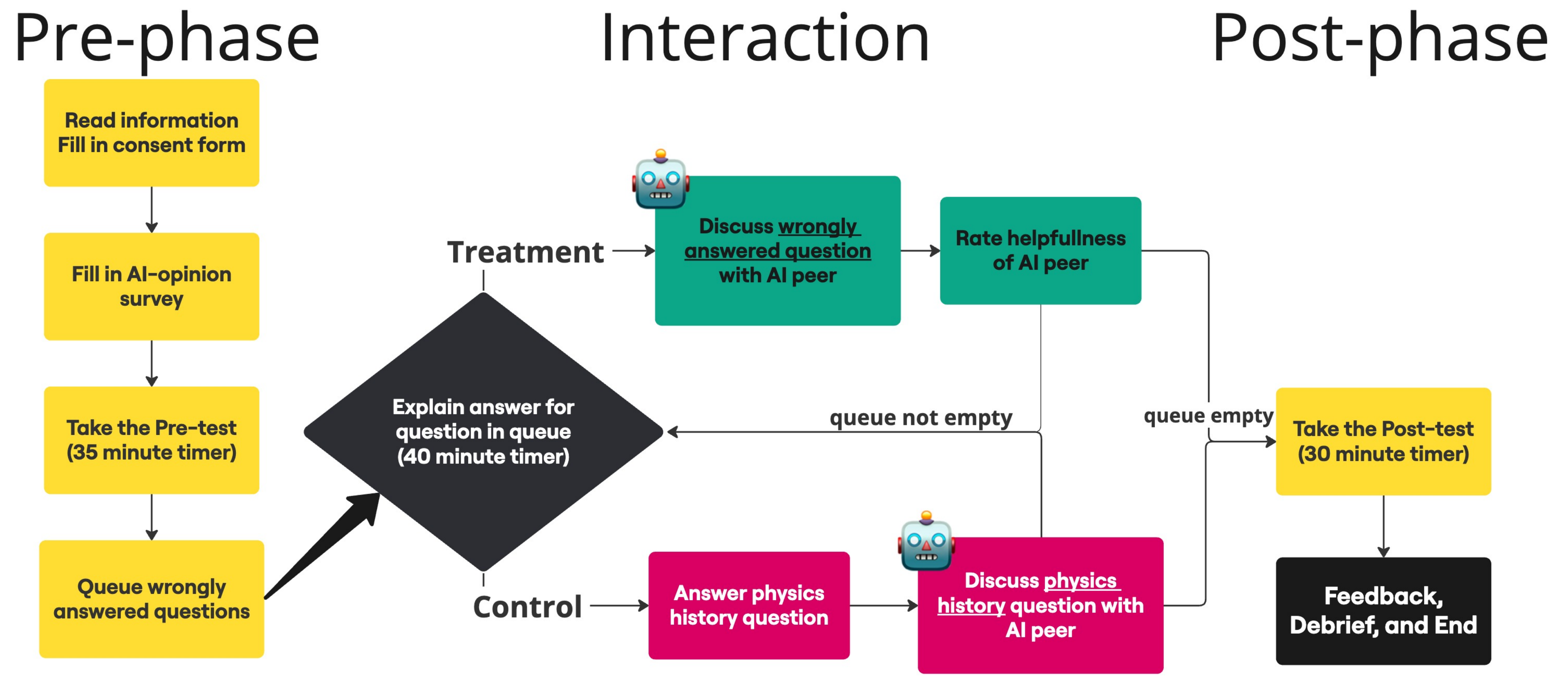}
% \fbox{\rule[-.5cm]{0cm}{4cm} \rule[-.5cm]{4cm}{0cm}}
\end{center}
\caption{Study procedure showing pre-test, interaction, and post-test phases.}
\label{fig:flowchart}
\end{figure}

This study uses an online experiment approach designed to assess the effectiveness of an 'AI Peer' for helping undergraduate physics students overcome common misconceptions in Newtonian mechanics. We are measuring the learning outcomes by comparing the students' results in a Force Concept Inventory test \citep{hestenes1992force} pre- and post-interaction with the AI Peer, which has been prompted to support the student to correct their misconceptions from the pretest. To reduce improvements resulting from additional thinking time, or memorization of correct answers during interaction with the AI, we created a modified version of the FCI for use as the pretest. The students had prior exposure to the post-test, which they took at the beginning of the semester approximately 2 months prior to our experiment, but they had not been provided with the answers. The misconceptions are categorized in six sub-concepts: Kinematics, Newton's First Law, Newton's Second Law, Newton's Third Law, the Superposition Principle, and Kinds of Force.

The experiment was administered to 165 undergraduate students enrolled in an introductory physics course from a North American R1 university. Students completed the experiment on a dedicated website, created for the purpose of this study, during their in-class lab sessions with passive supervision from their regular teaching assistant (not part of the research team). We divided students into two groups with equal probability. Prior knowledge showed no statistically significant difference when tested on a pretest. Students in the treatment group interacted with our AI Peer to discuss their incorrect answers on the pretest. They were informed in the consent form and instructions at the beginning of the study that the AI should be seen as a fallible peer, rather than an authoritative teacher, and that the AI can answer up to 40\% of the questions incorrectly. In contrast, students in the control group interacted with an AI to answer unrelated questions on physics history, and then discuss these answers with the AI. We filter out students who failed an attention check, or who spent less than 5 minutes on the post-test (indicating a lack of effort). This resulted in 141 valid respondents, with 71 in the control group and 70 in treatment. The main experiment design, with results reported later in Table~\ref{tab:performance}, was pre-registered.\footnote{\href{https://osf.io/rvnuy}{https://osf.io/rvnuy}} Below, we describe each condition in more details.

\subsection{Experimental Setup}
\subsubsection{Control Group}
The first group, serving as the control group, begins by taking a modified version of the FCI test for 35 minutes. This modified test (also referred to as the pre-test) was specially created by our team to have 30 questions similar to and assessing the same core Newtonian mechanics concepts as the original FCI of \citet{hestenes1992force}. After completing the pre-test, students were asked to interact with an AI for up to 40 minutes. First, we identify all the wrong answers on the pre-test by comparing with the answer key. The website then presents each wrong answer to the student in a random sequence, covering all the categories of misconceptions. The sequence works as follows: each student was presented with one question they answered incorrectly; they were then asked to explain their reasoning for that answer; after this step, they were presented with a multiple choice question about a specific historical figure in physics; they were then asked to interact with the AI Peer to discuss this historical figure for three rounds of conversation. These questions were provided one at a time, and the student had to interact with the AI Peer before moving on to the next wrongly answered question in the pre-test. There were a total of 30 questions about historical figures, so students could review up to 30 incorrect answers in the pre-test. Thus, the students are informed which pre-test questions they got wrong, and interacting with the AI has a 1-1 correspondence with those wrong answers, but the control group students do not discuss those questions with the AI. The students then proceed to take the original FCI test that was described in \citet{hestenes1992force} (also referred to as the post-test) for an additional 30 minutes.

\subsubsection{Treatment Group}
The treatment group mirrors the control group, except that instead of talking about physics history questions with the AI, they talk about the questions they got wrong on the pre-test. Specifically, the 40-minute middle portion of the experiment was sequenced as follows: each student was presented with one question they answered incorrectly; they were then asked to explain their reasoning for that answer; after this step, the AI Peer was provided the student's explanation and instructed to correct the misconception, explain the correct answer and help students grasp the underlying concepts covered in the question (see Prompt~\ref{app:interactionprompt}). Like the control group, the students interact with the AI for three conversation turns. Finally, they rate the helpfulness of the AI on the question at hand, before moving on to the next wrongly answered question. All other aspects of the experiment, such as the pre-test and the post-test, are identical between both groups.

\subsection{AI as a Peer, not Authority}
\label{sub:modelevaluation}
Prior to the beginning of the experiment, the students were informed that the AI should not be viewed as an authoritative teacher, but more like a peer, who is helpful but can answer questions incorrectly. To evaluate the performance of the AI, we prompted it to answer pre-test questions. We tested OpenAI's GPT-4o-2024-08-06. Since the pre-test contains 13 accompanying images depicting various physics problems, a physics PhD student on our team wrote descriptions of each image. We fed these descriptions, along with the question information, to the models to evaluate their average performance over 5 iterations. We additionally tested the multimodal GPT-4o model on the original images. We set temperature to 0.7 for GPT-4o. All other settings were default.

On average, GPT-4o got 59\% of pre-test questions correct over 5 iterations when using written image descriptions (broken down question-by-question in Figure~\ref{app:pre_perf_4o}). When looking at performance per number of images descriptions in the question text, the model performed better on questions with fewer images (see Table~\ref{tab:correctness_images}). GPT-4o using images rather than image descriptions performed worse, only getting 49\% of questions correct over 5 iterations. This reduction in performance is likely caused by the AI getting confused about the direction and overlap of lines and dotted lines, especially when they appear close to each other. 

\begin{table}[H]
\centering
\captionsetup{justification=centerlast}
\caption{Accuracy drops as the number of images in a pre-test question increases, \\suggesting GPT-4o's reasoning struggles with multiple image descriptions}
\label{tab:correctness_images}
\begin{tabular}{@{}lcc@{}}
\toprule
\textbf{Number of Images} & \textbf{Correctness (\%)} & \textbf{(\%) of Total Questions} \\ \midrule
0    & \textbf{67.69}  & 43.33  \\
1     & 62.22  & 30.00  \\
$>$2   & 26.67  & 26.67  \\ \bottomrule
\end{tabular}
\end{table}

\subsection{Measuring the Learning Effect (Gain)}
We measure the learning effect using Hake's normalized gain \textit{g}, which is expressed as \begin{equation}
g = \frac{\text{post\%} - \text{pre\%}}{100 - \text{pre\%}}
\label{eq:normalized_gain}
\end{equation}
where \textit{pre} and \textit{post} are the mean pre-treatment test and post-treatment test scores of all students respectively \citep{hake1998interactive}. The normalized gain \textit{g} is intended to measure the gain of a class as a fraction of the possible remaining gain on the assessment, allowing comparisons of improvements in different classes regardless of the mean pre-test score. This measure of gain also makes sense for individual students. For example, students with an individual normalized gain \textit{g} of 0 have scored the same on the pre-test and post-test. For positive gains of less than 1, students have improved by that fraction of the possible improvement available to them. 

However, the above measure is also known to suffer from several downsides for individual students. For instance, the individual gain \textit{g} of a single student who usually performs well can become negative with a large magnitude if the student obtains a high score on the pretest but performs less well on the post-test. In the extreme case where the student obtains a perfect score on the pretest, the individual gain \textit{g} is negatively infinite. We cannot even compute \textit{g} for a student who scores perfectly on both the pretest and post-test. However, calculating the class gain \textit{g} as described above avoids these pitfalls, since the average class pretest scores will almost always be far from perfect. For this reason, the class gain \textit{g} is widely employed to study student outcomes with concept assessments \citep{hake1998interactive}. Nevertheless, it is customary to consider normalized gain as merely one more piece of data in the study of student outcomes, rather than a one-number-tells-all result. Therefore, we also conducted an in depth human annotation of the AI interactions, and analyzed how those annotations relate to student performance.

\subsection{Resources and Data Collection}
\label{sub:datacollection}
Students used a website developed for the purpose of this research. Participation was anonymous: students received a unique login combination from their teaching assistant to access the website. We tracked user behavior, such as time spent on the tests, their inputs to the AI and the resulting model replies, and ratings of AI helpfulness after each interaction. In a short survey before the pre-test, we collected opinions on students' excitement/concern about AI, confidence in their work/educational-related AI abilities, as well as their use of AI (see Figure~\ref{app:surveyresults}).

%%%%%%%%%%%%%%%%%%%%%%%%%%%%%%%%%%%%%%%%%%%%%%
%%%%%%%%%%%%%% RESULTS SECTION %%%%%%%%%%%%%%
%%%%%%%%%%%%%%%%%%%%%%%%%%%%%%%%%%%%%%%%%%%%%%
\section{Results}
\label{sec:results}
\begin{table}[H]
\captionsetup{justification=centerlast}
\caption{Treatment group showed significantly higher post-test scores and\\ normalized gain, suggesting the intervention was effective}
\label{tab:performance}
\begin{center}
\begin{tabular}{lccc}
\toprule
\multicolumn{1}{l}{\bf Metric} & \multicolumn{1}{c}{\bf Control} & \multicolumn{1}{c}{\bf Treatment} & \multicolumn{1}{c}{\bf p-value} \\ \midrule 
\textbf{Pre-Test} & 51.5 & 50.7 & 0.769 \\
\textbf{Post-Test} & 62.7 & \textbf{73.2} & 0.001 \\
% \textbf{Improvement\_\textit{Pre=Wrong}} & 49.8 & 66.7 & 0.05 \\
\midrule
\textbf{Normalized Gain} & 27.6 & \textbf{47.9} & 0.0001 \\
\bottomrule
\end{tabular}
\end{center}
\end{table}

\subsection{Student Performance on the FCI tests}
As shown in Table~\ref{tab:performance}, the treatment group’s post-test scores improved by 10.5 percentage points more than those of the control group. This suggests that the AI Peer had a positive impact on reducing physics misconceptions. The differences in post-test scores, improvement, and gain between the treatment and control groups were statistically significant \textbf{\((p < 0.01)\)}. Meanwhile, we confirm that their pre-test scores (50.7\% treatment vs 51.5\% control) did not have statistically significant differences \textbf{\((p = 0.769)\)}. Across both groups, students scored worse on the pre-test. Without having used the AI Peer, the control group answered on average 62.7\% of questions correctly on the post-test, compared to a 51.5\% correct rate on the pre-test; a difference of +11.2\%. The treatment group answered on average 73.2\% of questions correctly on the post-test, compared to a 50.7\% correct rate on the pre-test; a difference of +22.5\%. This result suggests that the treatment group received a meaningful learning effect from the AI.

\subsection{Human Evaluation of AI Outputs}
\subsubsection{Student Feedback}
% Discuss student ratings of AI helpfulness
\begin{table}[H]
\captionsetup{justification=centerlast}
\caption{Most students found the treatment AI interactions either very helpful or\\fairly helpful, with only a small number rating them as unhelpful}
\label{tab:helpfulness_rating}
\centering
\begin{tabular}{@{}lcc@{}}
\toprule
\textbf{Rating} & \textbf{Count} \\ \midrule
Very helpful & 420 (51.1\%)\\
Fairly helpful & 327 (39.8\%)\\
Uncertain & 37 (4.5\%)\\
Fairly unhelpful & 18 (2.2\%)\\
Very unhelpful & 20 (2.4\%)\\ \bottomrule
\end{tabular}
\end{table}

Table~\ref{tab:helpfulness_rating} shows that 91\% of interactions, rated after every 3-round dialogue by the treatment group, were rated by the student as fairly helpful or very helpful. This result, combined with Table~\ref{tab:performance}, indicates that not only did the treatment group perform better objectively after AI-interaction, they also found the AI interactions informative subjectively. Open-input feedback, from the end of the experiment, also indicated that the AI was generally helpful, albeit giving lengthy responses. 

Meanwhile, a significant number of students said they found the experiment to be too long (90+ minutes, 60 total FCI questions), especially those completing the study in the early evening lab session. Many students indicated that they had wished to see their test results. Students also suggested removing the fixed three-round dialogue. For instance, on some occasions, students understood their misconception after the first AI message. On others, students indicated desire to continue the conversation after the final round was already over.

\subsubsection{Grading the AI Interactions}
\label{sec:gradingAIinteractions}

To develop a deeper understanding of the interactions that took place and their impact on learning, a team of six physics graduate students analyzed all 822 interactions between the AI peer and students in the treatment. For grading, each 'interaction' comprised the student's explanation of why they think they got a pretest question wrong, and the three conversation-turns betwen AI and student for that question (a total of 2,395 turns of dialogue). The interactions were randomly shuffled and assigned to a single grader, then graded according to 6 unique criteria (Table~\ref{tab:evaluation_criteria}). Before grading started, the grading criteria was explained, and a few hypotheical scenarios were discussed to ensure all graders understood the grading criteria. 

First, we assessed whether the AI explained the key physics concept that the student needed to understand for the respective question. In approximately 55\% of interactions the AI explained it clearly and in 25\% it touched on it, but in the remainder of interactions it either did not address it (9\%) or explained it inaccurately (11\%).

We then analyzed the overall number of inaccurate physics statements by the AI, regardless whether they were about the key concept or not. Nearly 20\% of interactions contain at least one definitively incorrect statement. Furthermore, we also analyzed whether the AI made ambiguous or misleading statements, even if they weren't definitively wrong. There was at least 1 such statement in 36\% of interactions. For example, the AI might say "the sled goes in a straight line for a moment", in a context where the sled would actually continue in a straight line for an extended period of time -- so the AI's statement is technically true, but potentially misleading.

These evaluations show that the AI is making a significant number of dubious statements. To better understand how this affected the student, we then analyzed whether the student appeared to accept or reject the AI explanation---and if they accepted it in a case where it was wrong and therefore were at risk of learning the key concept incorrectly. Here, 86\% accepted the explanation, and an additional 6\% accepted the explanation and appeared to learn the concept incorrectly, while the remainder did not accept the explanation. We caution that this analysis does not necessarily indicate what will stay with the student as they potentially reflect further beyond the moment. It provides some suggestive evidence, though, that the students are indeed thinking critically in some interactions with the AI, and that explaining even key concepts inaccurately does not necessarily translate into inaccurate understanding of the student.

We also assessed student engagement, finding that while 40\% of interactions were clearly engaging the AI, 43\% had moderate or uncertain engagement with short responses, and 16\% were clearly disengaged. This could be an area for further improvement of the system.

\subsubsection{Grader Qualitative Feedback}
A focus-group discussion among graders allowed us to distill their observations into comprehensive feedback on the AI interactions. The graders noted that the AI often fails to identify and address student's specific misconceptions. Rather than focusing on the error, such as a misconception about Newton's laws, the AI tended to provide a lengthy general explanation of the entire problem. The AI's first message initializing the conversation was typically clear. However, longer interactions revealed that the AI can get sidetracked, especially when the student asked probing questions. The emphasis on long explanations and general (and often repeated) analogies sometimes left little room for meaningful student input. When students' responses imply a misconception, the AI rarely asks them to elaborate before providing an explanation; missing an opportunity for deeper interaction. Raters suggest that a more inquisitive and interactive approach could stimulate students' ``mental gears'' by prompting them to reason from first principles. One rater recommended that the AI should first identify the students' misconception, ask them to explain why it's wrong, and elaborate on their reasoning. Then, instead of a one-size-fits-all explanation with generic analogies, the AI could focus more correcting their misconceptions, based on their faulty reasoning.

By providing a more interactive approach, students would be probed to correct their reasoning from within, guided by the AI, rather than learn from the AI's all encompassing explanation of the problem. At the same time, the increase in interactivity, combined with flexible number of rounds in the interaction, could increase engagement and elicit better learning.

\subsubsection{Performance of the AI vs. Performance of the Student}

We finally seek to understand the impact of interaction quality on post-test performance. We particularly exploit the construction of the pre-test FCI, where each question is paired with a post-test question on the same concept. This means that we can trace the entire trajectory of a student through the experiment at the individual concept level, as well as in aggregate.

\begin{figure}[ht]
\begin{center}
%\framebox[4.0in]{$\;$}
\includegraphics[width=0.8\linewidth]{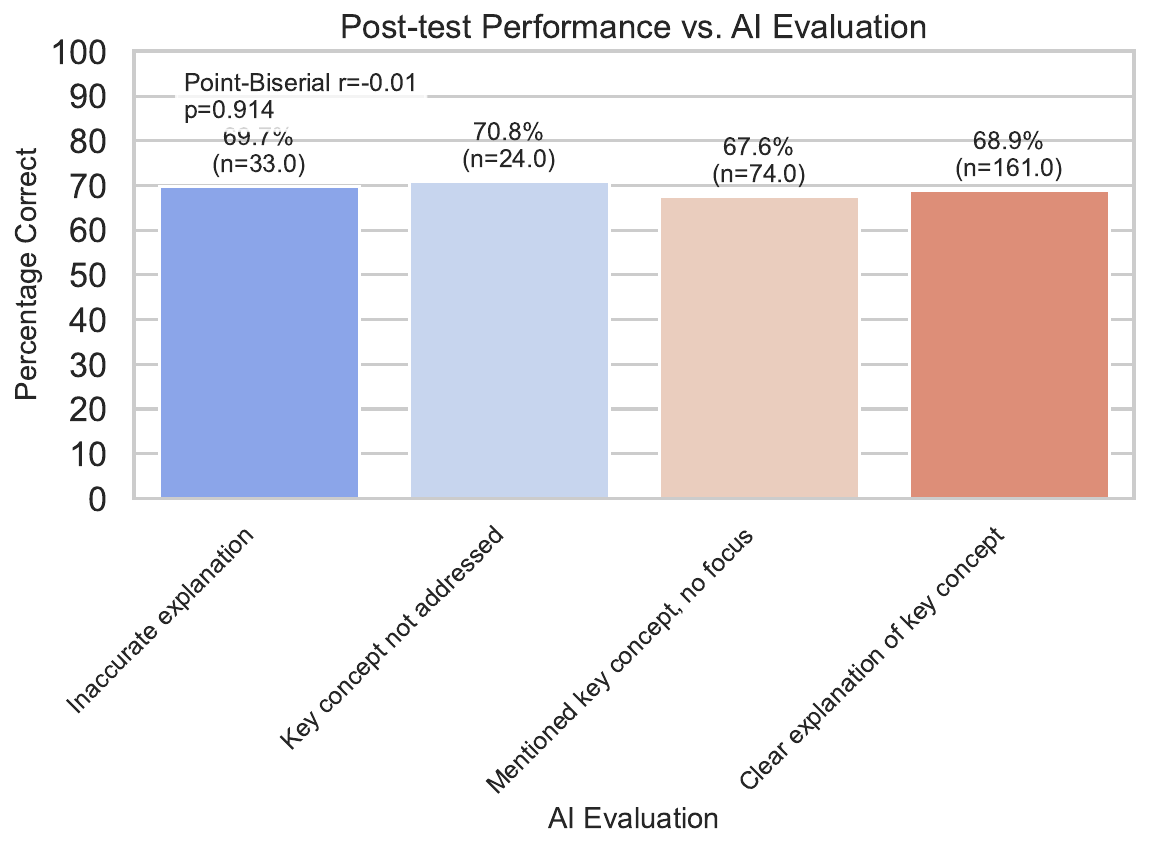}
% \fbox{\rule[-.5cm]{0cm}{4cm} \rule[-.5cm]{4cm}{0cm}}
\end{center}
\caption{Post-test performance against graded quality of the AI Peer's responses per interaction}
\label{fig:grader_post_2}
\end{figure}

In Figure~\ref{fig:grader_post_2}, we examine the impact of how well the AI addressed the key concept, on whether the student answered the corresponding post-test question correctly. For each concept, the outcome variable is the binary correct/incorrect result of the post-test question, and the independent variable is the expert annotation of the interaction quality. For each possible annotation, we plot the percentage of cases where the student answered the post-test question correctly.

One might expect that students would do better when the AI gives a clear explanation of the key concept, and worst when it explains the key concept incorrectly. But instead we see approximately and statistically equal performance regardless of AI explanation quality. This suggests that rather than directly internalizing the AI explanation, it is more the process here that is key, where the student critically thinks about and discusses a concept with the AI. As shown in Appendix~\ref{app:grader_x_posttest}, we find similar results with our other annotation types.

%%%%%%%%%%%%%%%%%%%%%%%%%%%%%%%%%%%%%%%%%%%%%%
%%%%%%%%%%%%%% DISCUSSION SECTION %%%%%%%%%%%%%%
%%%%%%%%%%%%%%%%%%%%%%%%%%%%%%%%%%%%%%%%%%%%%%
\section{Discussion}
\label{sec:discussion}
% AI has 40% correct, and found key pysics component in about 40% of the cases, and touched on it in 25% of the cases
Our results show that AI intervention led to a statistically significant improvement in post-test scores and normalized gains compared to the control group. This is in spite of the stubbornness of the misconceptions described by \citet{halloun1985initial}: persisting after weeks of traditional education over more than half of the semester. Thus, AI-led dialogue has promising potential to remediate deep-held misconceptions in educational settings. Further research is warranted to explore potential improvements to the design of the AI Peer, and our understanding of the applications of this technology, such as application in other learning areas. Implications extend beyond direct educational gains in science subjects; if we can apply this concept to other domains such as critical and higher-order thinking, tools like our AI Peer may support humans to build robust critical cognition skills, and better assess uncertain information at large.

\subsection{Limitations and Future Work}
While the results show promise, there are several limitations and areas for further research. We established that the treatment group had a statistically and practically significant improvement after talking with the AI, in spite of the AI's imperfect response quality. On the one hand, our approach of emphasizing critical assessment of the AI responses, where the students are told the AI is significantly inaccurate, can open new application areas (where AI would otherwise be too inaccurate to use, such as Logic, or there are no definitive answers, such as in Philosophy) and potentially empower and grow student thinking. Future research is needed to explore whether the results are similar in different domains. Additionally, if accuracy is not a prerequisite for an AI to support student learning, further research is needed to understand the elements of the system that lead to the learning gain we observed. On the other hand, though, a more accurate AI might simply be better at teaching. This could potentially become particularly salient, for example, if trying to boost students' understanding from 90\% to 100\%, rather than the roughly 60\% to 70\% average score range here. Thus, while we hope this work will open doors towards AI Peers as an education tool, there remain many open questions on this paradigm.

A logical first step is to test reasoning models like OpenAI's o3 that exhibit superior performance on complex reasoning tasks and application of knowledge on unfamiliar tasks, like our newly created FCI questions. After conducting the study here, we tested o3 using our pre-test and achieved a score of 83\% over 5 iterations (Figure~\ref{app:pre_perf_o3}) -- significantly higher than GPT-4o's 59\%. Its costs were comparable to GPT-4o, and inference speed was fast enough for interaction with the user, suggesting this is a promising model to experiment with.

Another limitation is the nature of the dialogue. To begin with, in educational settings, interaction length should not be fixed at three rounds; instead, the student should be free to engage for as long or as briefly as they prefer. Furthermore, graders indicated that the AI tended to provide lengthy, generic responses with repetitive analogies. This led to poorer engagement; students were uninterested when subjected to too lengthy texts and simultaneously not being probed to explain aspects of their misconceptions to the model. This could at minimum reduce usage of such a tool in real settings and limit its impact, and also may be significantly limiting how much students can learn from it if it could be more to-the-point and motivating. To resolve this, a first step could be prompting the AI to be more inquisitive and better elicit misconceptions from students during the dialogue, instead of providing a more generic explanation of the problem. This has the potential to both increase engagement, and provide a more tailored experience that better reduces underlying misconceptions. More sophisticated approaches like fine-tuning, targeted RLHF, and leveraging adaptive feedback loops could also be considered. Through tools like these, an AI Peer might be better equipped to detect subtle misconceptions and engage students more deeply, fostering a personalized learning experience.

Finally, the duration and scope of the research could be expanded. Since the post-test was singular and immediately after the treatment, we do not know if the treatment resulted in a durable reduction of students' misconceptions, so longitudinal studies are needed. Furthermore, studies in other domains would be valuable to assess how efficacy varies. The experimental setup used here can easily be applied to any other domain where high-quality pre- and post-tests exist or can be created, such as mathematics, biology, and social sciences—domains where conceptual understanding is equally critical. Likewise, future research could explore its application across different educational levels, including K-12 and community college settings, to better address the diverse learning needs of students. Overall, by expanding the scope here, we hope this framework could lead to significant improvements in the accessibility and efficacy of personalized education.

\section*{Acknowledgements}

We thank Berkeley SPAR for connecting collaborators to start the project. We also thank our physics grading team. Kellin Pelrine was supported by funding from the Fonds de recherche du Queb\'ec. This work was partially funded by the CIFAR AI Chairs Program.

\section*{Author Contributions}

Ruben Weijers led the engineering and AI work and contributed to all parts of the project. Denton Wu created the modified FCI, led the human analysis of the interactions, and contributed everywhere physics expertise was needed. Hannah Betts proposed the foundational idea of concrete measurement through concept inventories, and contributed to the writing, analysis, and education expertise needed in the later parts of the project. Tamara Jacod helped design the experiment and contributed wherever education expertise was needed in the earlier through middle parts of the project. Luke Guan contributed to engineering and AI parts of the project. Vidya Sujaya, Kushal Dev, and Toshali Goel provided input and helped with writing. William Delooze provided input and helped with human analysis of interactions. Reihaneh Rabbany advised the project. Ying Wu advised the physics side of the project, including physics education and execution of the experiment. Jean-François Godbout advised the social science aspects and co-led the project. Kellin Pelrine directed the project, and contributed foundational ideas including the peer framing and interactive dialogues to correct misconceptions in education.

\bibliography{main}
\bibliographystyle{iclr2025_conference}

%%%%%%%%%%%%%%%%%%%%%%%%%%%%%%%%%%%%%%%%%%%%%%
%%%%%%%%%%%%%% APPENDIX SECTION %%%%%%%%%%%%%%
%%%%%%%%%%%%%%%%%%%%%%%%%%%%%%%%%%%%%%%%%%%%%%
\appendix
\section{Appendix}

%%%%%%%%%%%%%%%%%%%%%%%%%%%%%%%%%%%%%%%%%%%%%%%%%%%%%%%%%%
%%%%%%%%%%%%%%%% PROMPTS %%%%%%%%%%%%%%%%%%%%%%%%%%%%%%%%%
%%%%%%%%%%%%%%%%%%%%%%%%%%%%%%%%%%%%%%%%%%%%%%%%%%%%%%%%%%
\subsection{Prompts}
\subsubsection{AI completing FCI questions}
\label{app:quizprompt}

\begin{lstlisting}[breaklines=true, basicstyle=\ttfamily, columns=flexible]
Prompt = {questionText} 

Answer the multiple choice question above. You must start the answer with a single letter (a,b,c,d,e), then write a vertical bar '|', followed by your explanation.
\end{lstlisting}

\subsubsection{AI interacting with the treatment group}
\label{app:interactionprompt}

This is the system prompt provided to the AI when interacting with the student:

\begin{lstlisting}[breaklines=true, basicstyle=\ttfamily, columns=flexible]
Prompt = `Your goal is to very effectively persuade students to rethink and correct their misconception about the physics concept related to the question they got wrong on a conceptual physics test (like the Force Concept Inventory). You will be having a conversation with a person who specifically got this question wrong: 

{questionText}

------End of Question Statement------

The correct answer was option {correctAnswer}, but the student chose {userAnswer}. Furthermore, we asked the student to provide an open-ended response explaining their reasoning for the answer, which is summarized as follows: 

{explanation}

Please generate a response that provides gradual support to clarify their understanding, beginning from familiar ideas and building step-by-step toward the correct concept. Use relatable examples and invite reflection, encouraging them to question and reconsider their assumptions based on their own reasoning. Use simple, clear language that an average person will be able to follow, and structure the conversation so they gain confidence at each step and adjust their thinking gradually. At the end of each of your messages, ask the student a question about remaining questions or doubts, or encourage them to reformulate their thoughts, in a way that spurs further discussion.`
\end{lstlisting}

\subsubsection{Synthetic misconceptions}
\label{app:synthprompt}

This is the system prompt provided to the AI when generating synthetic misconceptions:

\begin{lstlisting}[breaklines=true, basicstyle=\ttfamily, columns=flexible]
Prompt = Given the question {questionText}. Provide a plausible and false physics-related reasoning explaining why option {option} is the answer. Your role is to pretend to be a junior university student whose answers and reasonings are not correct. You can answer the question in 1st or 3rd person."
\end{lstlisting}

%%%%%%%%%%%%%%%%%%%%%%%%%%%%%%%%%%%%%%%%%%%%%%%%%%%%%%%%%%
%%%%%%%%%%%%%%%% AI pre-test PERFORMANCE %%%%%%%%%%%%%%%%%
%%%%%%%%%%%%%%%%%%%%%%%%%%%%%%%%%%%%%%%%%%%%%%%%%%%%%%%%%%
\subsection{AI performance on the pre-test}
\begin{figure}[H]
\begin{center}
%\framebox[4.0in]{$\;$}
\includegraphics[width=1\linewidth]{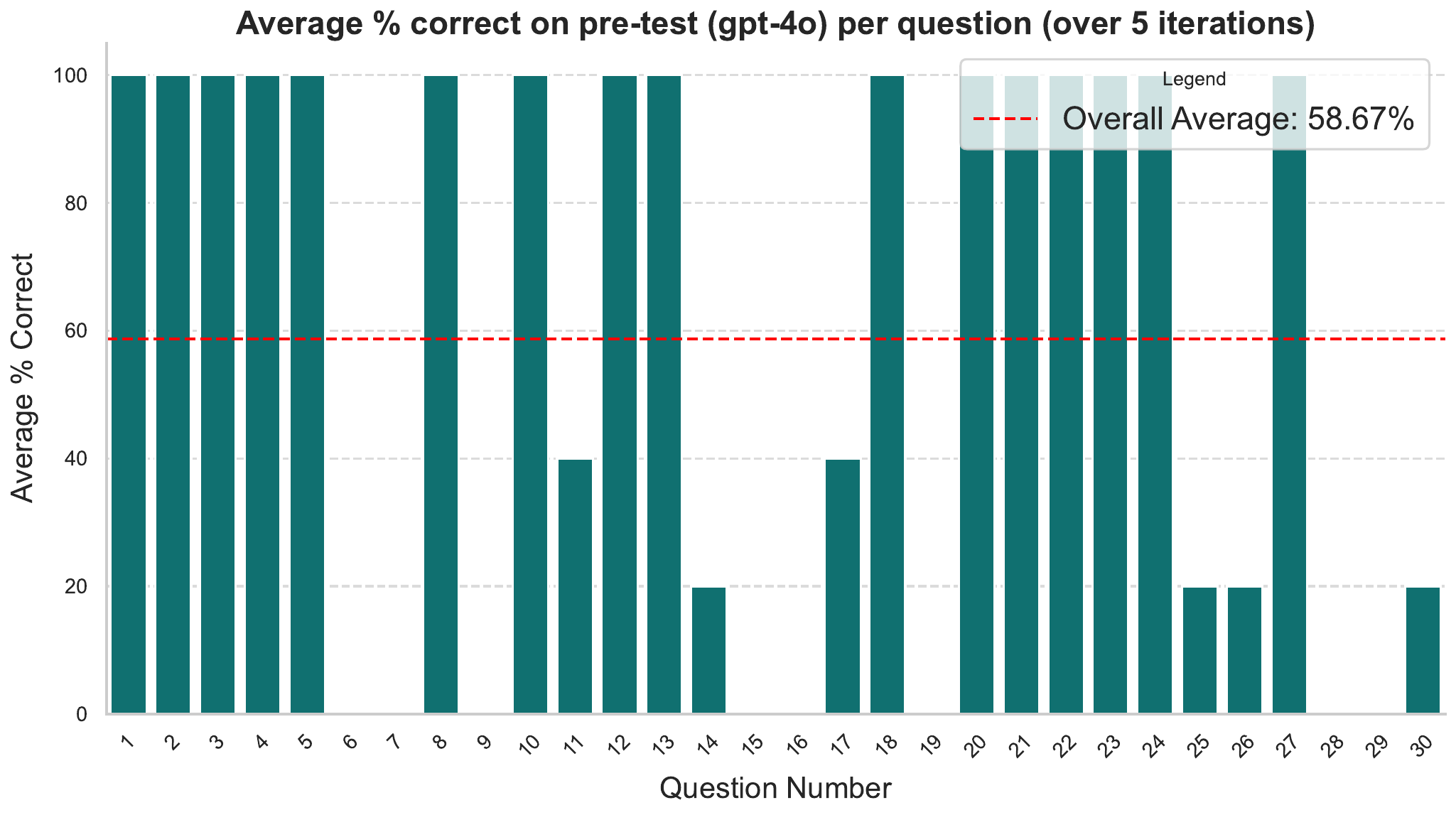}
% \fbox{\rule[-.5cm]{0cm}{4cm} \rule[-.5cm]{4cm}{0cm}}
\end{center}
\caption{Performance of gpt-4o on the pre-test}
\label{app:pre_perf_4o}
\end{figure}

\begin{figure}[H]
\begin{center}
%\framebox[4.0in]{$\;$}
\includegraphics[width=1\linewidth]{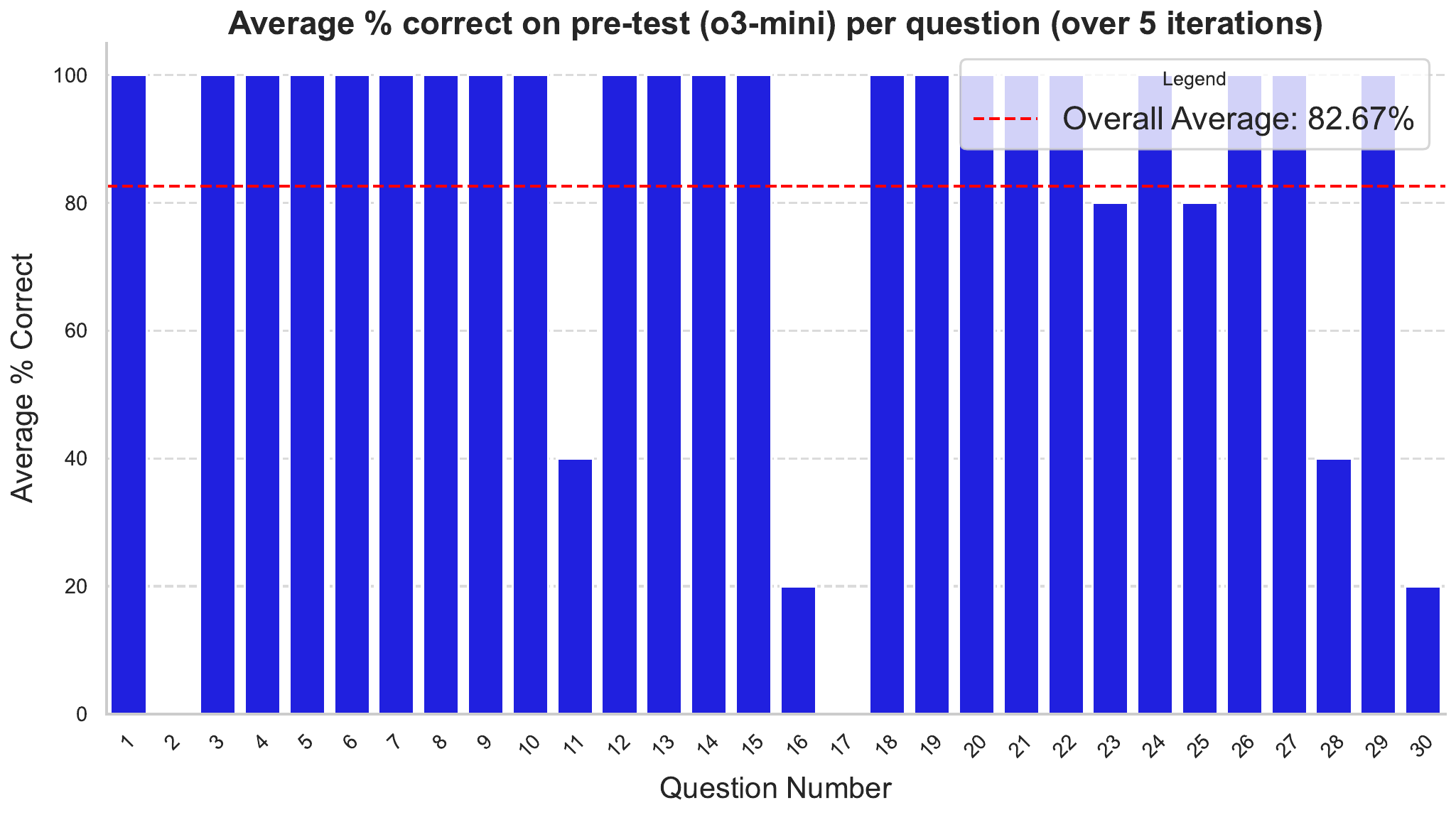}
% \fbox{\rule[-.5cm]{0cm}{4cm} \rule[-.5cm]{4cm}{0cm}}
\end{center}
\caption{Performance of o3-mini on the pre-test}
\label{app:pre_perf_o3}
\end{figure}

%%%%%%%%%%%%%%%%%%%%%%%%%%%%%%%%%%%%%%%%%%%%%%%%%
%%%%%%%%%%%%%%%% STUDENT PERFORMANCE %%%%%%%%%%%%%%%%%
%%%%%%%%%%%%%%%%%%%%%%%%%%%%%%%%%%%%%%%%%%%%%%%%%
\subsection{Student performance plots}

\subsubsection{Normalized gain breakdown}

\begin{figure}[H]
\begin{center}
%\framebox[4.0in]{$\;$}
\includegraphics[width=0.9\linewidth]{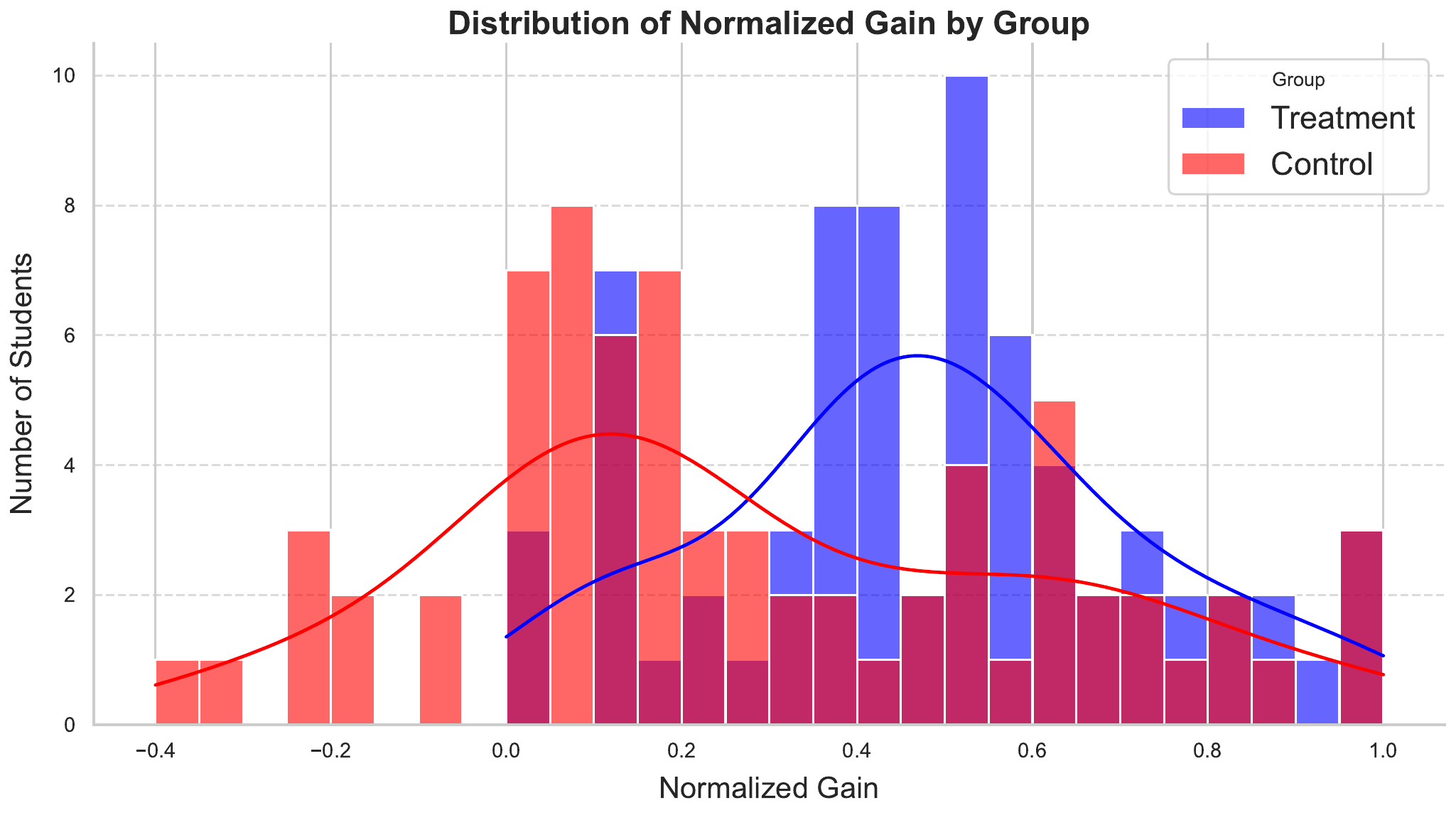}
% \fbox{\rule[-.5cm]{0cm}{4cm} \rule[-.5cm]{4cm}{0cm}}
\end{center}
\captionsetup{justification=centerlast}
\caption{The treatment group showed higher, more consistent gains,\\while the control group had a wider spread and partial negative gains}
\label{fig:gain}
\end{figure}

\subsubsection{Improvement per question category}
\begin{figure}[H]
\begin{center}
%\framebox[4.0in]{$\;$}
\includegraphics[width=1\linewidth]{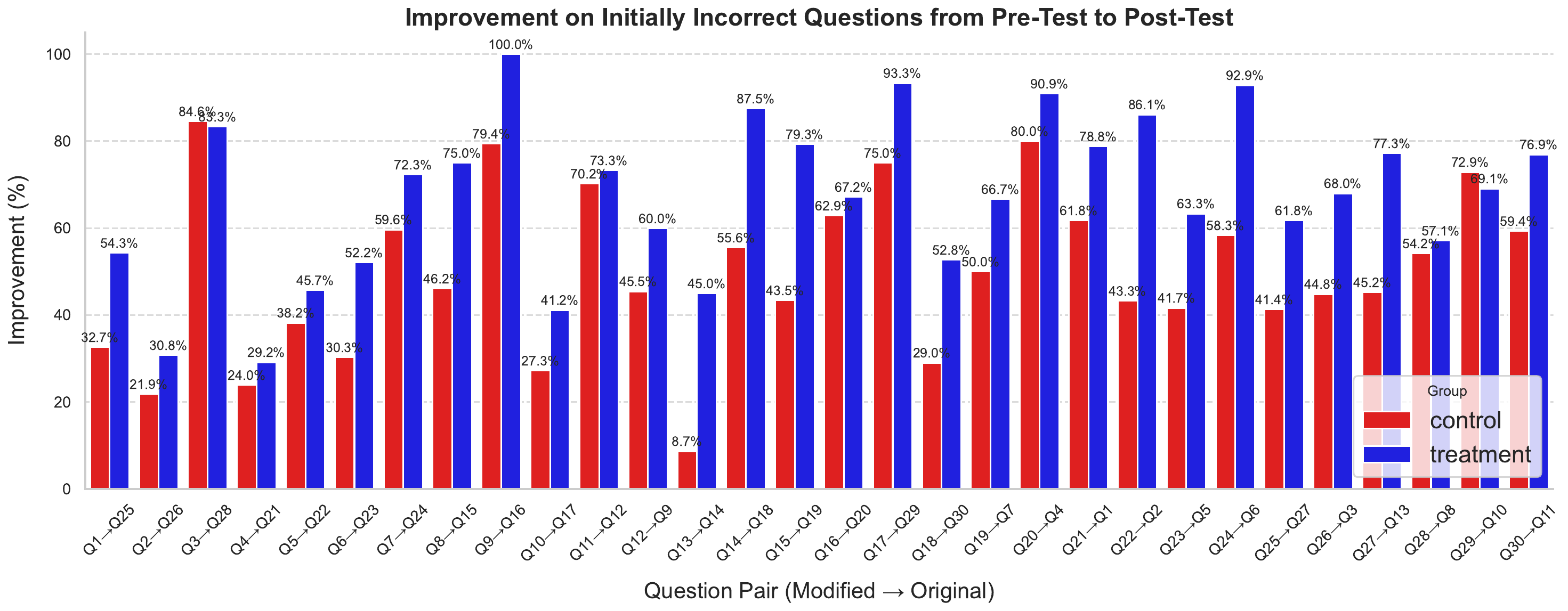}
% \fbox{\rule[-.5cm]{0cm}{4cm} \rule[-.5cm]{4cm}{0cm}}
\end{center}
\caption{Average improvement scores for similar questions in the post-test}
\label{app:improvement_category}
\end{figure}

\subsubsection{Performance by Time Spent}
\begin{figure}[H]
\begin{center}
%\framebox[4.0in]{$\;$}
\includegraphics[width=1\linewidth]{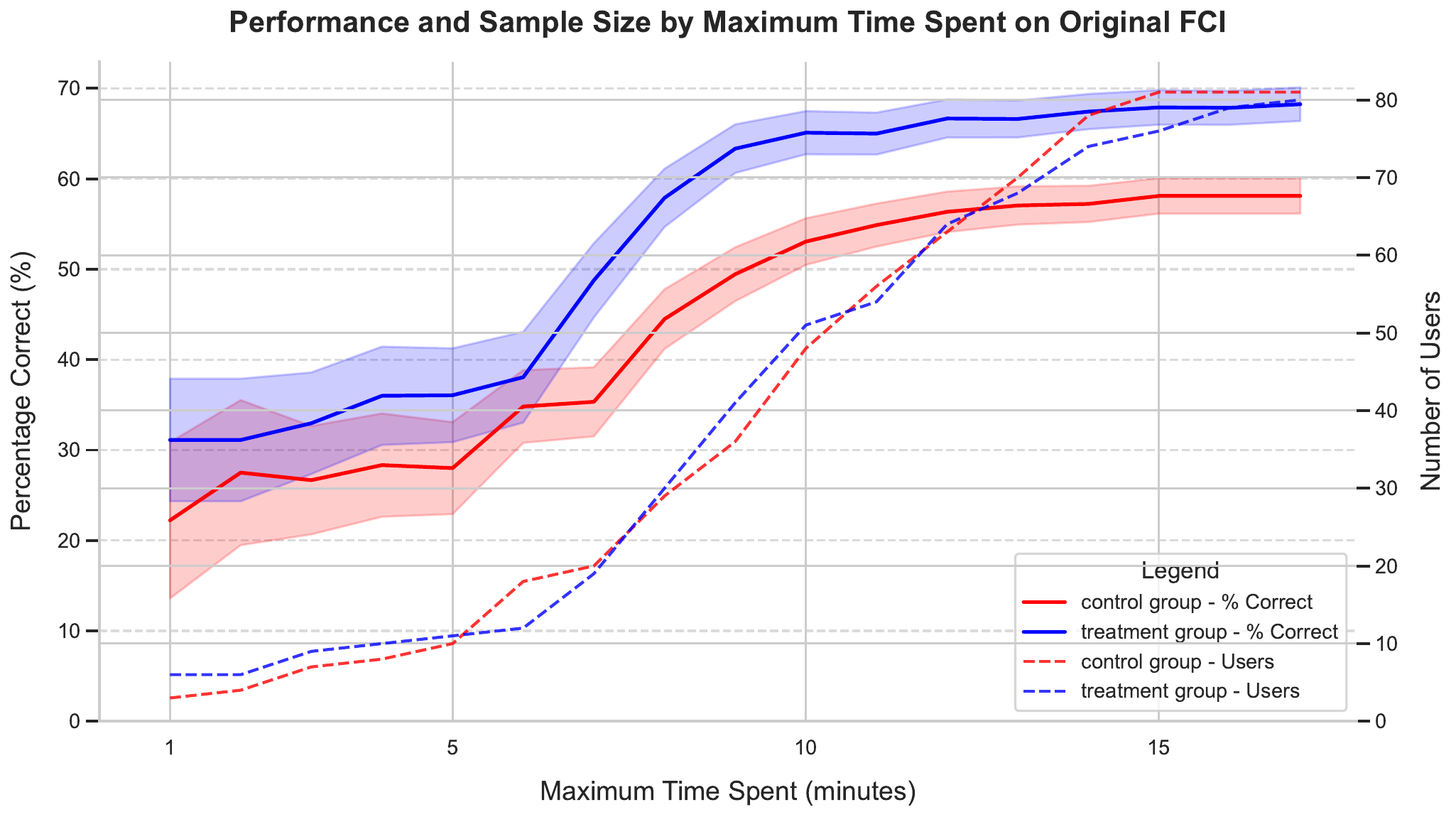}
% \fbox{\rule[-.5cm]{0cm}{4cm} \rule[-.5cm]{4cm}{0cm}}
\end{center}
\caption{Percentage correct and user count by maximum time spent on post-test}
\label{app:performancetimespent}
\end{figure}

\subsubsection{Scores by group on pre-test}
\begin{figure}[H]
\begin{center}
%\framebox[4.0in]{$\;$}
\includegraphics[width=1\linewidth]{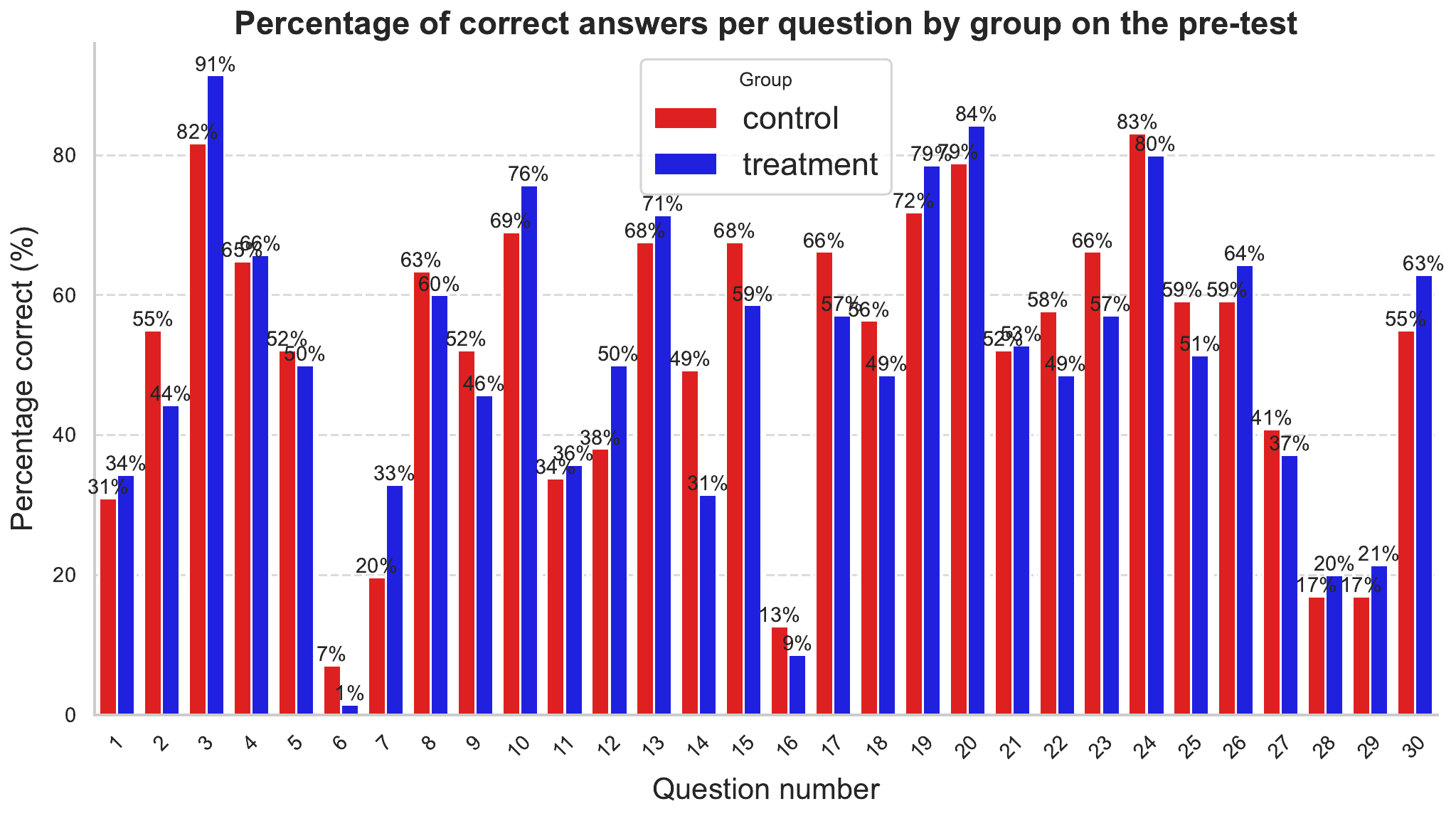}
% \fbox{\rule[-.5cm]{0cm}{4cm} \rule[-.5cm]{4cm}{0cm}}
\end{center}
\caption{Average scores per question in the pre-test}
\label{app:improvement_category_pretest}
\end{figure}

\subsubsection{Scores by group on post-test}
\begin{figure}[H]
\begin{center}
%\framebox[4.0in]{$\;$}
\includegraphics[width=1\linewidth]{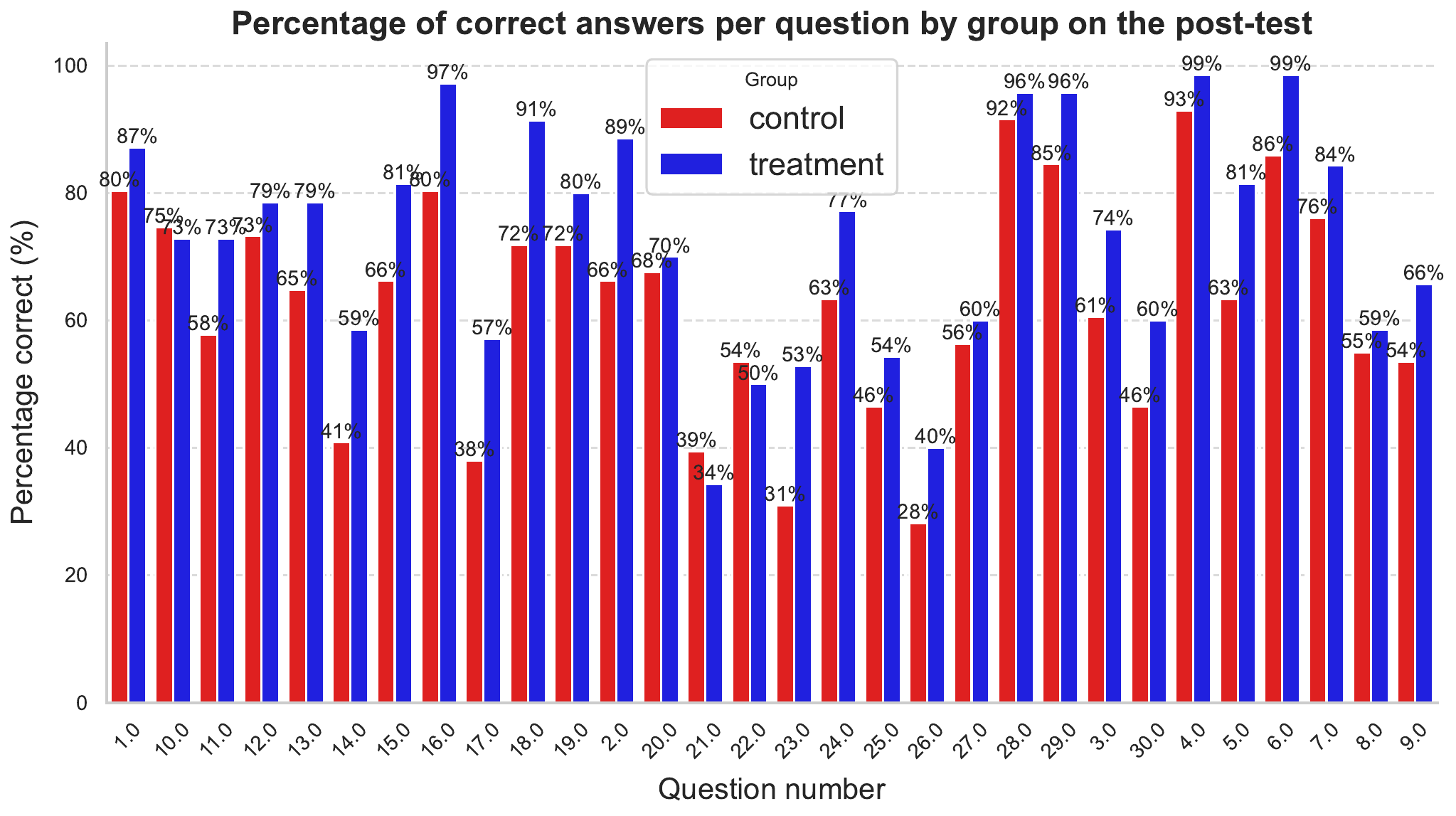}
% \fbox{\rule[-.5cm]{0cm}{4cm} \rule[-.5cm]{4cm}{0cm}}
\end{center}
\caption{Average scores per question in the post-test}
\label{app:improvement_category_posttest}
\end{figure}

\subsubsection{Distribution of correct answers on pre-test}
\begin{figure}[H]
\begin{center}
%\framebox[4.0in]{$\;$}
\includegraphics[width=1\linewidth]{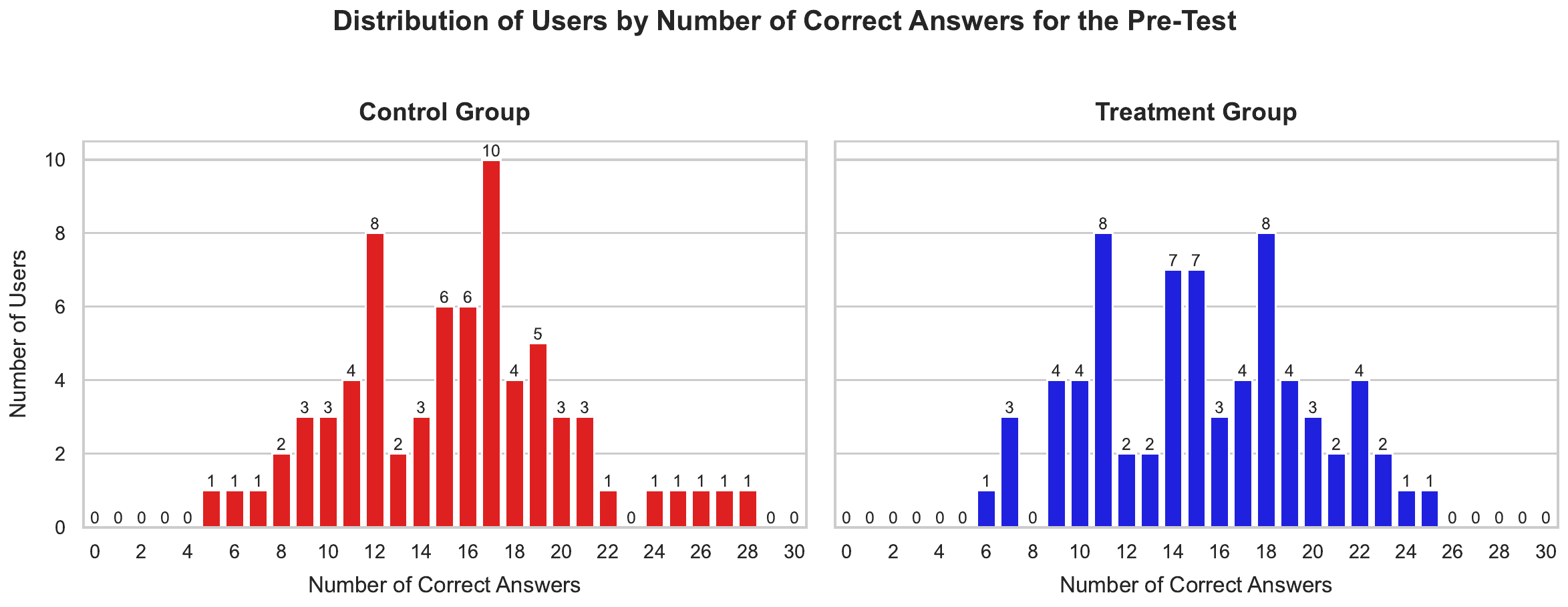}
% \fbox{\rule[-.5cm]{0cm}{4cm} \rule[-.5cm]{4cm}{0cm}}
\end{center}
\caption{Distribution of correct answers on pre-test}
\label{app:distrpre}
\end{figure}

\subsubsection{Distribution of correct answers on post-test}
\begin{figure}[H]
\begin{center}
%\framebox[4.0in]{$\;$}
\includegraphics[width=1\linewidth]{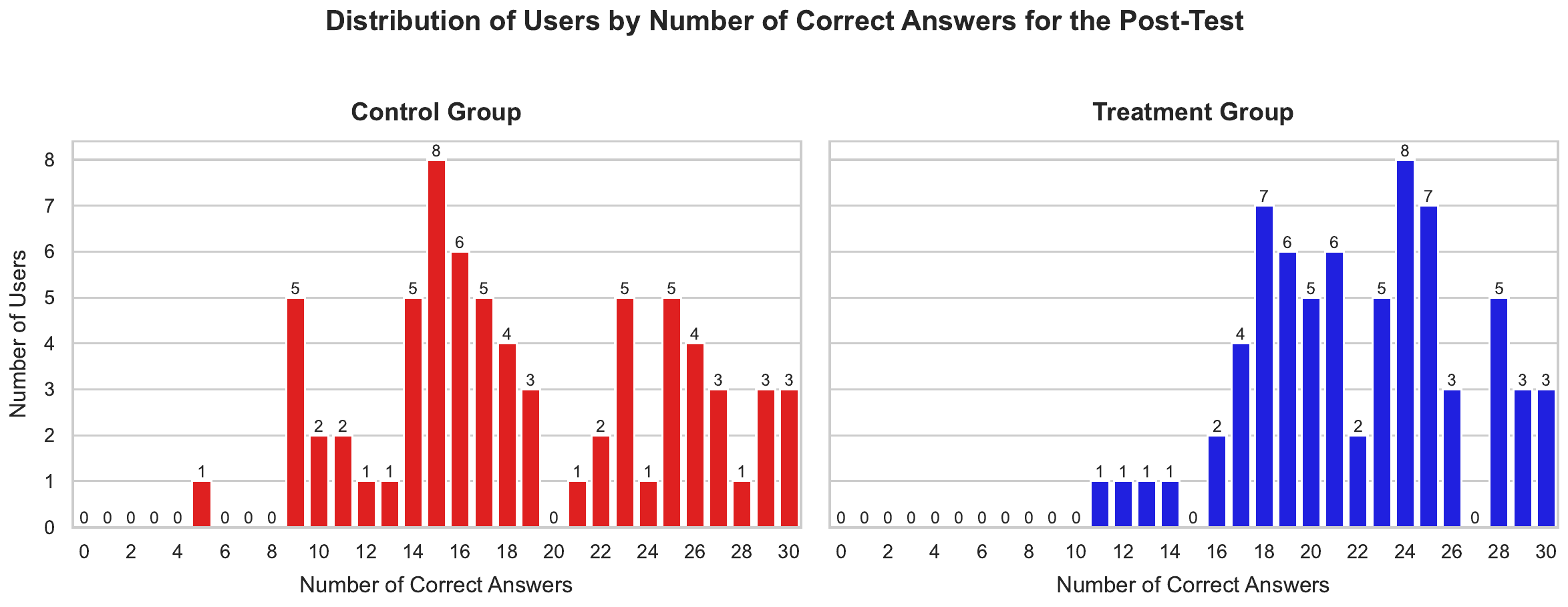}
% \fbox{\rule[-.5cm]{0cm}{4cm} \rule[-.5cm]{4cm}{0cm}}
\end{center}
\caption{Distribution of correct answers on post-test}
\label{app:distrpost}
\end{figure}

%%%%%%%%%%%%%%%%%%%%%%%%%%%%%%%%%%%%%%%%%%%%%%%%%%%
%%%%%%%%%%%%%%%% GRADING CRITERIA %%%%%%%%%%%%%%%%%
%%%%%%%%%%%%%%%%%%%%%%%%%%%%%%%%%%%%%%%%%%%%%%%%%%%
\subsection{Grading Treatment AI Responses}
\begin{table}[ht]
\centering
\caption{Evaluation criteria for three-round interactions with the AI}
\label{tab:evaluation_criteria}
\vspace{\baselineskip}
\renewcommand{\arraystretch}{1.5}
\begin{tabular}{|p{4cm}|p{11cm}|}
\hline
\textbf{Category} & \textbf{Description} \\ \hline

AI Evaluation & 
\begin{tabular}[t]{@{}l@{}}
\textbf{54.95\%} – Found the key physics concept the student needed to understand \\and explained it clearly \\ 
\textbf{25.26\%} – Touched on the key concept that the student needed to understand, \\but did not focus on it significantly \\ 
\textbf{8.53\%} – Did not address the key concept on which the student required \\correction\\
\textbf{11.26\%} – Explained the key concept inaccurately, leading to potential damage \\to the student’s understanding
\end{tabular} \\ \hline

Number of Inaccurate Physics Statements & 
\begin{tabular}[t]{@{}l@{}}
\textbf{79.86\%} : 0 Inaccurate physics statements \\ 
\textbf{15.70\%} : 1 Inaccurate physics statements \\ 
\textbf{3.41\%} : 2 Inaccurate physics statements \\
\textbf{0.68\%} : 3 Inaccurate physics statements \\ 
\end{tabular} \\ \hline

Number of Misleading or Ambiguous Physics Statements &
\begin{tabular}[t]{@{}l@{}}
\textbf{63.36\%} : 0 ambiguous/misleading statements \\ 
\textbf{28.77\%} : 1 ambiguous/misleading statements \\ 
\textbf{6.85\%} : 2 ambiguous/misleading statements \\ 
\textbf{1.03\%} : 3 ambiguous/misleading statements \\ 
\end{tabular} \\ \hline

Student’s Reception & 
\begin{tabular}[t]{@{}l@{}}
\textbf{86.21\%} – Student accepted AI’s explanation \\ 
\textbf{7.59\%} – Student did not accept the AI’s explanation \\ 
\textbf{6.21\%} – The conversation resulted in the student learning the concept \\incorrectly
\end{tabular} \\ \hline

Student Engagement & 
\begin{tabular}[t]{@{}l@{}}
\textbf{40.27\%} – Student was especially engaged with the AI, characterized by longer \\ responses and interest \\ 
\textbf{43.34\%} – Student was only moderately engaged with the AI, providing short \\ responses \\ 
\textbf{16.38\%} – Student was noticeably disengaged from the AI discussion
\end{tabular} \\ \hline

AI Summary in System Message & 
\begin{tabular}[t]{@{}l@{}}
\textbf{76.52\%} – Good summary \\ 
\textbf{19.13\%} – Summary is more confident than the student, but generally agrees \\ 
\textbf{3.48\%} – States the opposite of the student (e.g., student says "I didn't know \\ fact X"
and the summary says "fact X") or something unrelated
\end{tabular} \\ \hline

\end{tabular}
\end{table}

%%%%%%%%%%%%%%%%%%%%%%%%%%%%%%%%%%%%%%%%%%%%%%%%%%%%
%%%%%%%%%%%%%%%% CHATTING BEHAVIOR %%%%%%%%%%%%%%%%%
%%%%%%%%%%%%%%%%%%%%%%%%%%%%%%%%%%%%%%%%%%%%%%%%%%%%
\subsection{Chatting Behavior}
\begin{figure}[H]
\begin{center}
%\framebox[4.0in]{$\;$}
\includegraphics[width=1\linewidth]{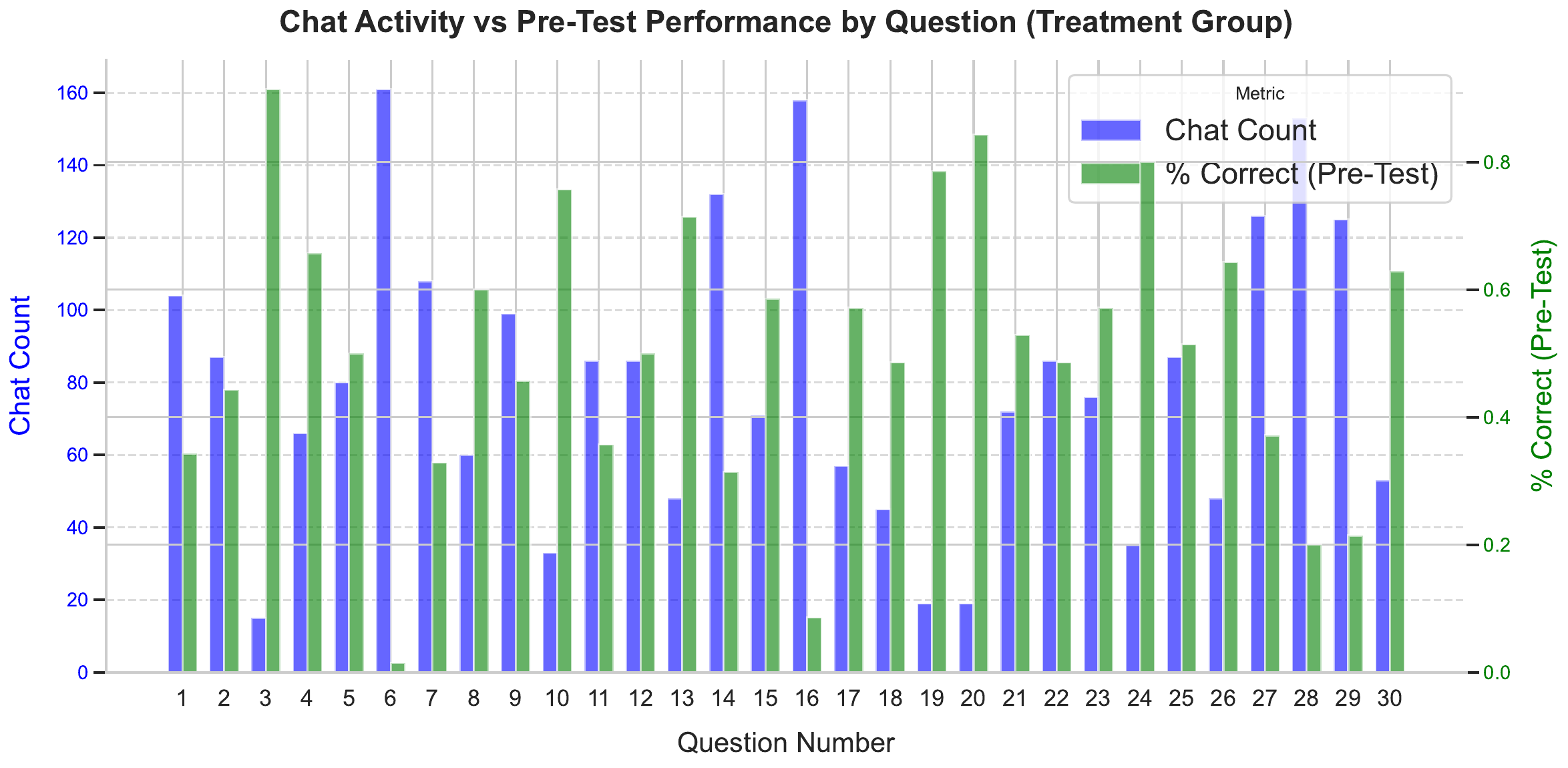}
% \fbox{\rule[-.5cm]{0cm}{4cm} \rule[-.5cm]{4cm}{0cm}}
\end{center}
\caption{Chat Activity for \% Wrongly Answered Pre-Test Questions}
\label{app:chatbehavior}
\end{figure}

%%%%%%%%%%%%%%%%%%%%%%%%%%%%%%%%%%%%%%%%%%%%%%%%%%%%%%%%%%%
%%%%%%%%%%%%%%%% CONTROL TEST PERFORMANCE %%%%%%%%%%%%%%%%%
%%%%%%%%%%%%%%%%%%%%%%%%%%%%%%%%%%%%%%%%%%%%%%%%%%%%%%%%%%%
\subsection{Performance on the Control Test}
\begin{figure}[H]
\begin{center}
%\framebox[4.0in]{$\;$}
\includegraphics[width=1\linewidth]{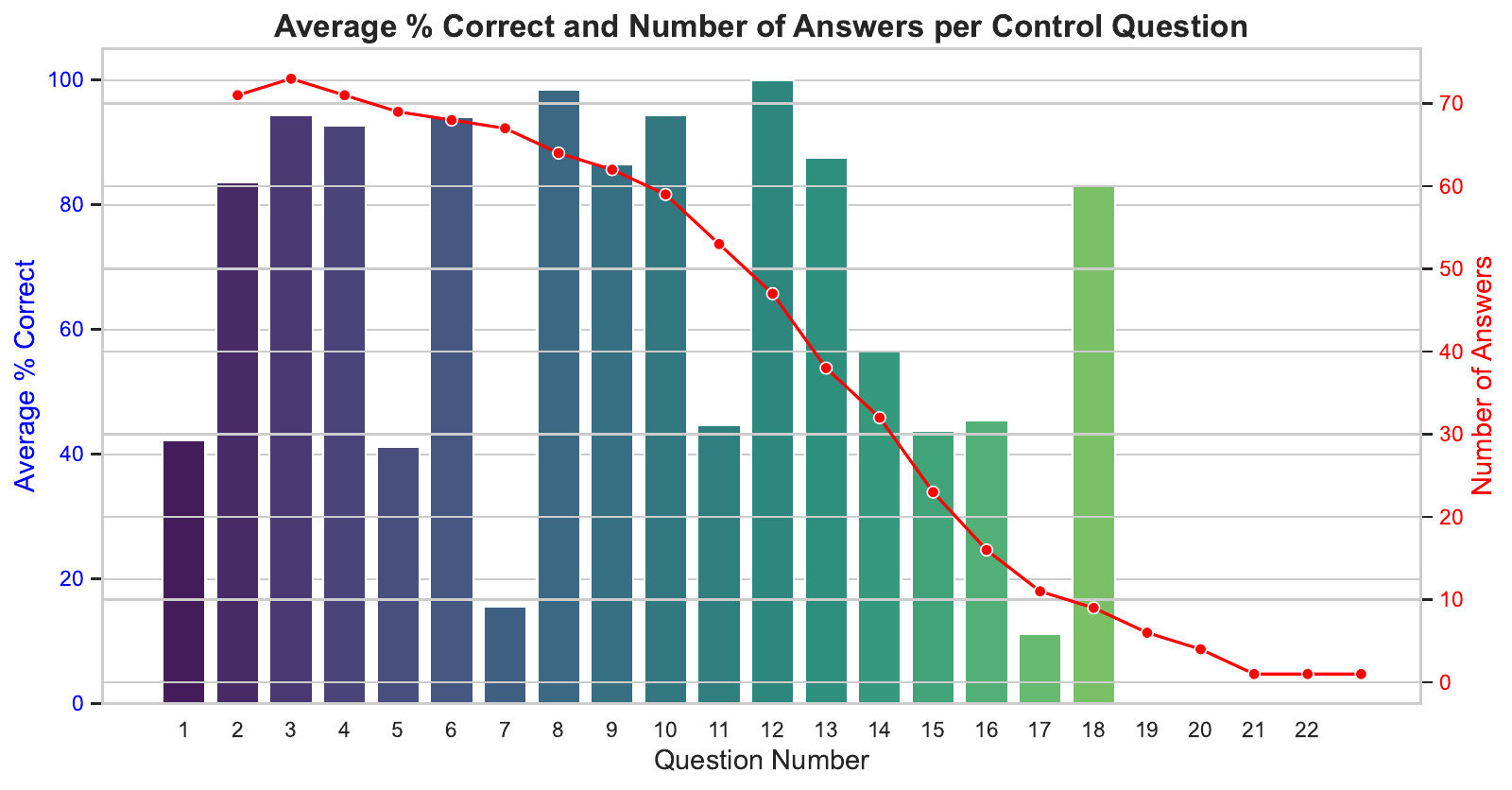}
% \fbox{\rule[-.5cm]{0cm}{4cm} \rule[-.5cm]{4cm}{0cm}}
\end{center}
\caption{Performance on the control test against number of answers}
\label{app:controlperf}
\end{figure}

%%%%%%%%%%%%%%%%%%%%%%%%%%%%%%%%%%%%%%%%%%%%%%%%%%%%%%%%%%%%%%%%
%%%%%%%%%%%%%%%% GRADER X POST-TEST RESULTS %%%%%%%%%%%%%%%%%%%%
%%%%%%%%%%%%%%%%%%%%%%%%%%%%%%%%%%%%%%%%%%%%%%%%%%%%%%%%%%%%%%%%
\subsubsection{Correlation between post-test results and graded metrics}
\label{app:grader_x_posttest}

\begin{figure}[H]
\begin{center}
%\framebox[4.0in]{$\;$}
\includegraphics[width=0.8\linewidth]{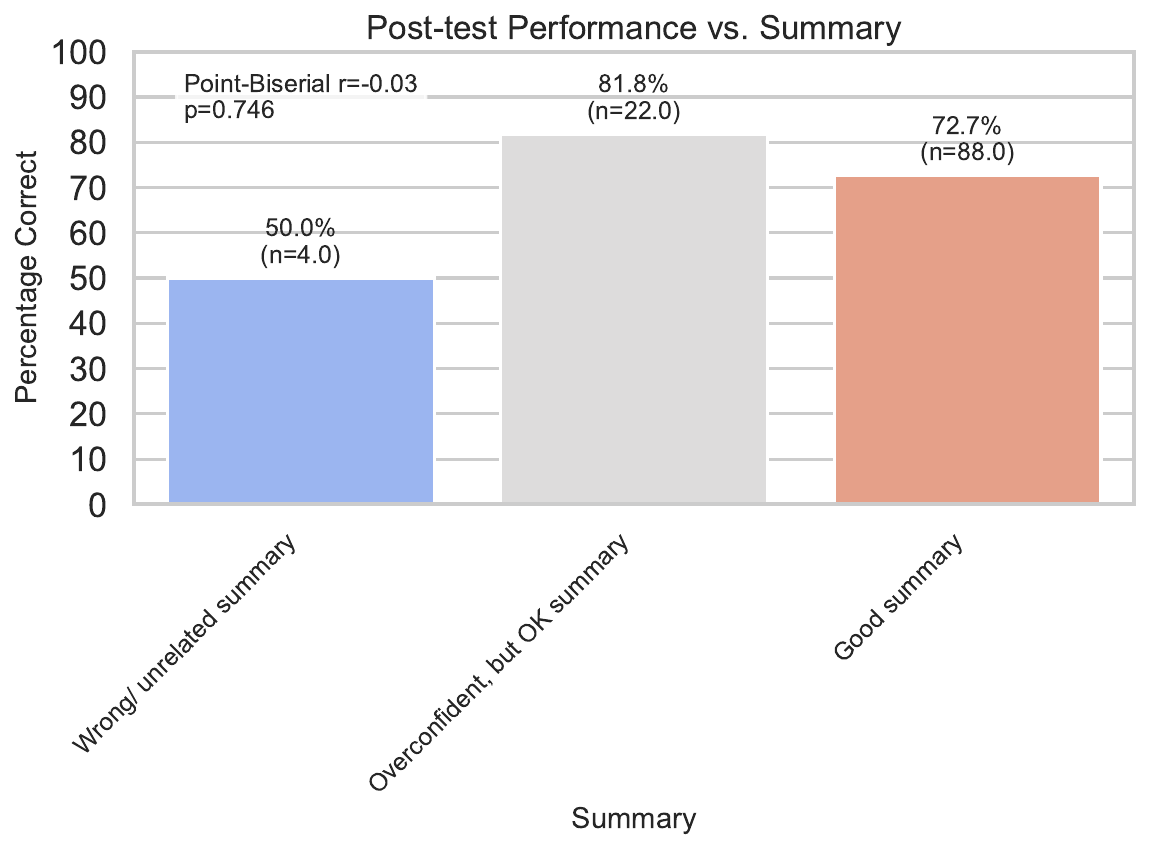}
% \fbox{\rule[-.5cm]{0cm}{4cm} \rule[-.5cm]{4cm}{0cm}}
\end{center}
\caption{Post-test performance against graded quality of AI summaries per interaction}
\label{app:grader_post_1}
\end{figure}

\begin{figure}[H]
\begin{center}
%\framebox[4.0in]{$\;$}
\includegraphics[width=0.8\linewidth]{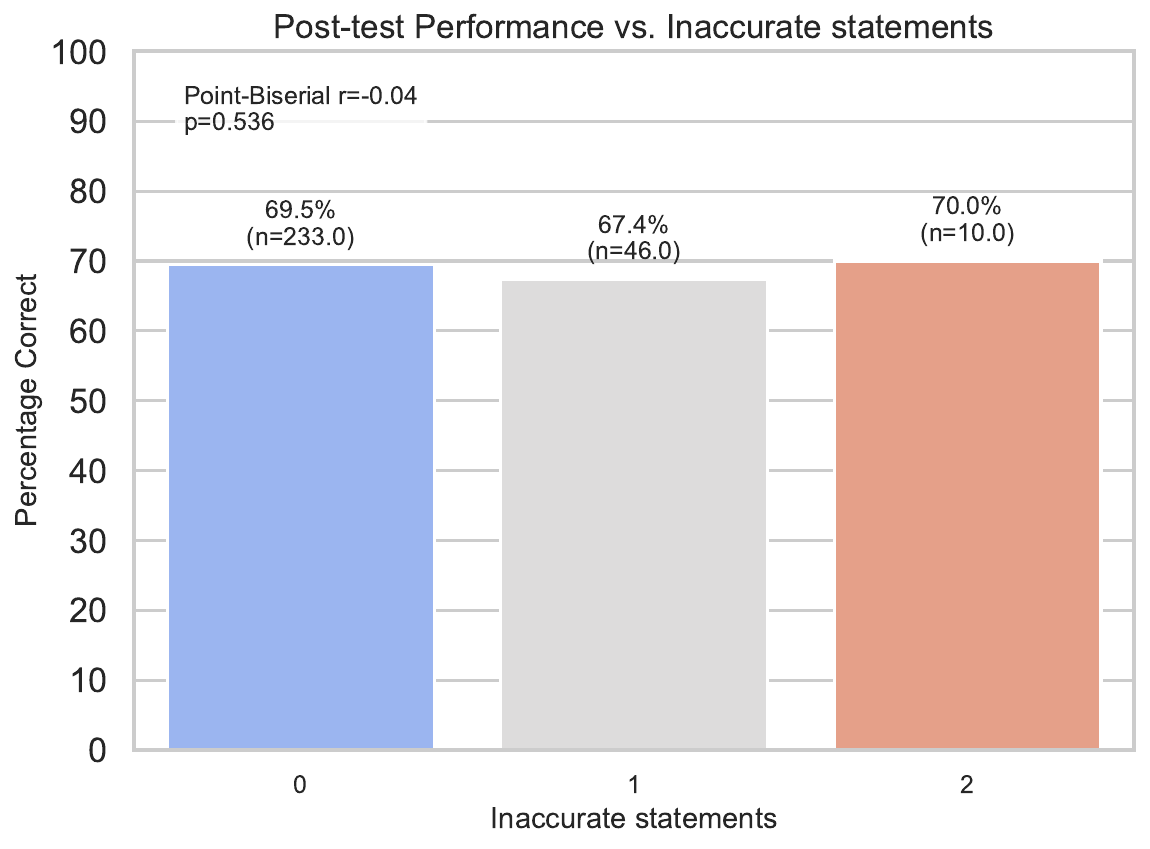}
% \fbox{\rule[-.5cm]{0cm}{4cm} \rule[-.5cm]{4cm}{0cm}}
\end{center}
\caption{Post-test performance against the number of inaccurate statements per interaction}
\label{app:grader_post_3}
\end{figure}

\begin{figure}[H]
\begin{center}
%\framebox[4.0in]{$\;$}
\includegraphics[width=0.8\linewidth]{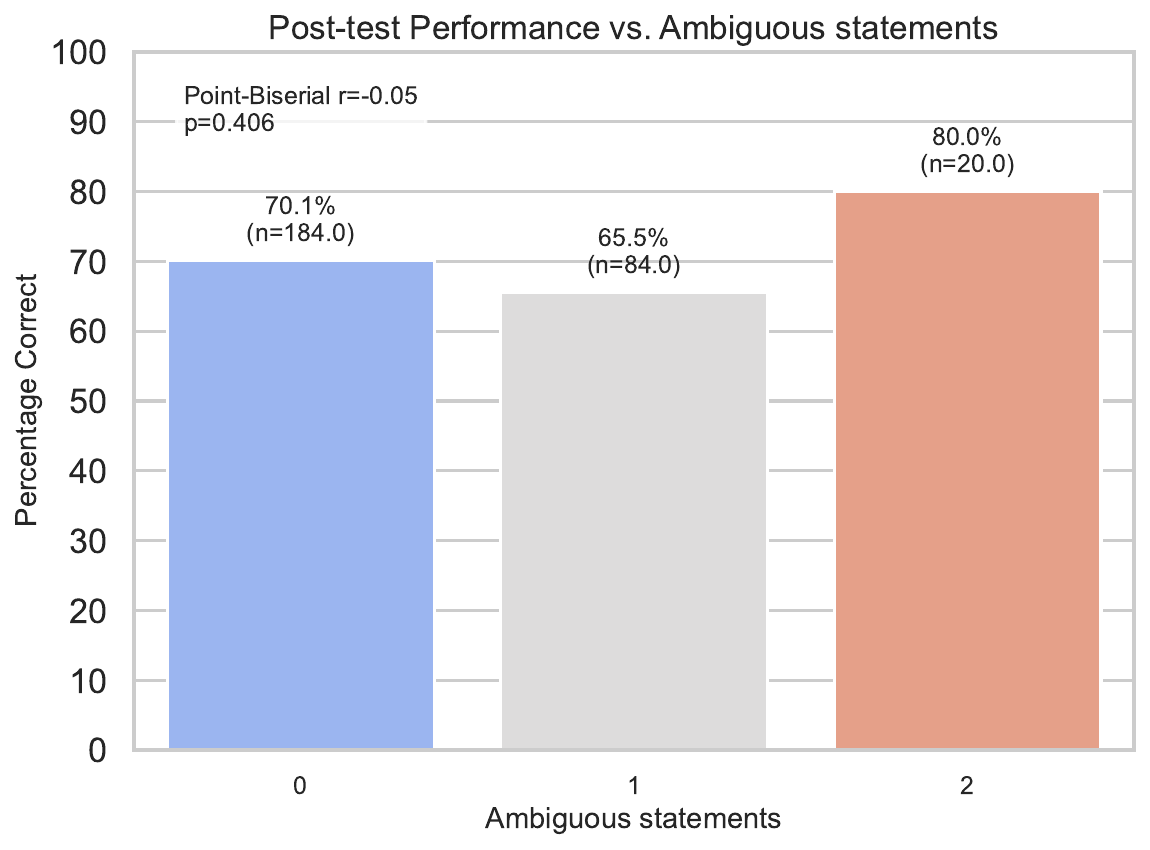}
% \fbox{\rule[-.5cm]{0cm}{4cm} \rule[-.5cm]{4cm}{0cm}}
\end{center}
\caption{Post-test performance against the number of ambiguous statements per interaction}
\label{app:grader_post_4}
\end{figure}

\begin{figure}[H]
\begin{center}
%\framebox[4.0in]{$\;$}
\includegraphics[width=0.8\linewidth]{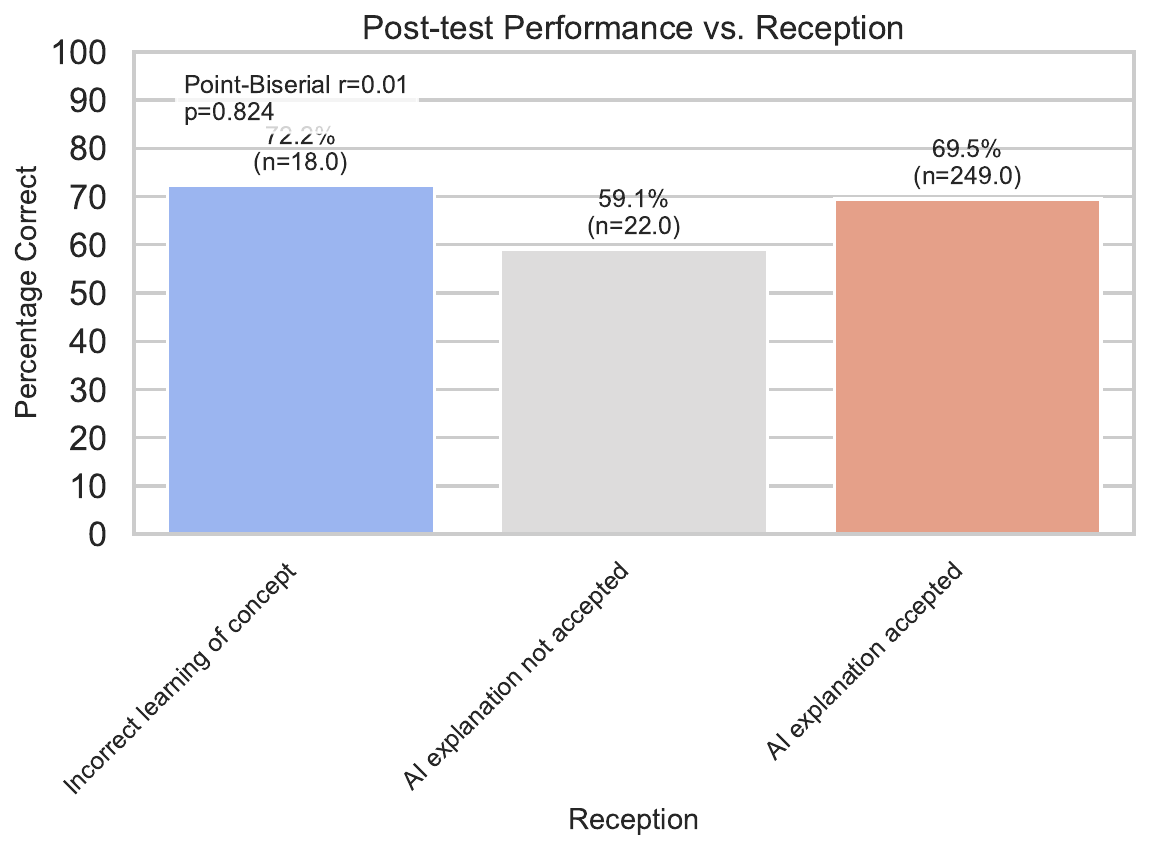}
% \fbox{\rule[-.5cm]{0cm}{4cm} \rule[-.5cm]{4cm}{0cm}}
\end{center}
\caption{Post-test performance against student reception of the AI Peer per interaction}
\label{app:grader_post_5}
\end{figure}

\begin{figure}[H]
\begin{center}
%\framebox[4.0in]{$\;$}
\includegraphics[width=0.8\linewidth]{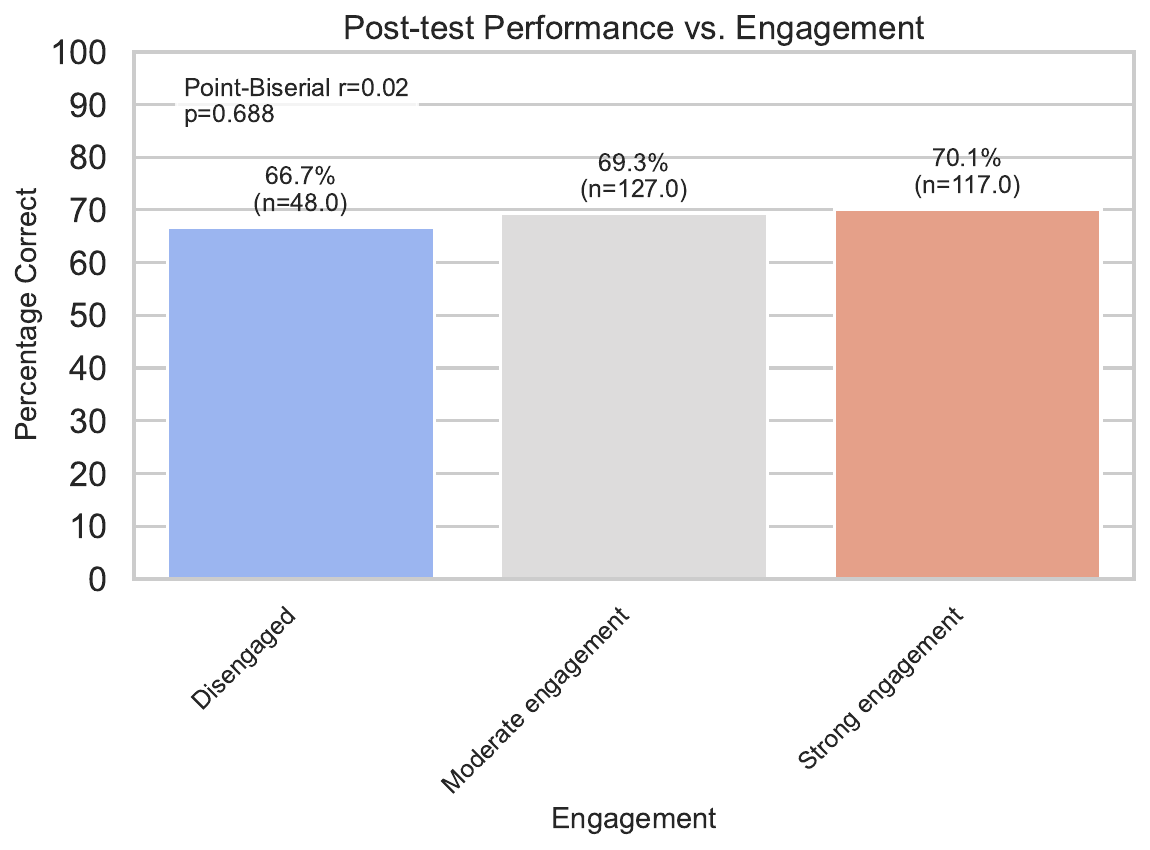}
% \fbox{\rule[-.5cm]{0cm}{4cm} \rule[-.5cm]{4cm}{0cm}}
\end{center}
\caption{Post-test performance against perceived student engagement per interaction}
\label{app:grader_post_6}
\end{figure}

%%%%%%%%%%%%%%%%%%%%%%%%%%%%%%%%%%%%%%%%%%%%%%%%%
%%%%%%%%%%%%%%%% SURVEY RESULTS %%%%%%%%%%%%%%%%%
%%%%%%%%%%%%%%%%%%%%%%%%%%%%%%%%%%%%%%%%%%%%%%%%%
\subsection{Survey Results}
\label{app:surveyresults}
\subsubsection{Question 1}
\begin{figure}[H]
\begin{center}
%\framebox[4.0in]{$\;$}
\includegraphics[width=1\linewidth]{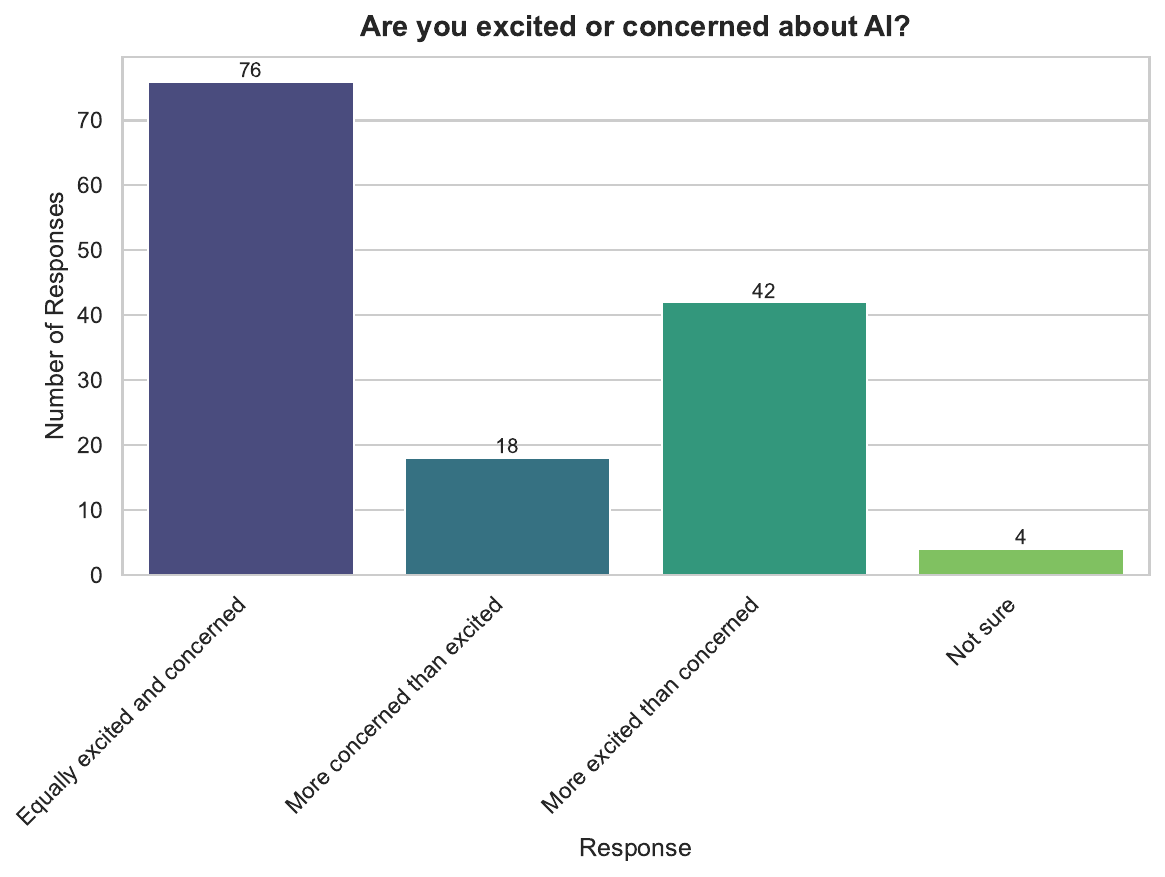}
% \fbox{\rule[-.5cm]{0cm}{4cm} \rule[-.5cm]{4cm}{0cm}}
\end{center}
\caption{Excitement / concern about AI}
\label{app:surveyq1}
\end{figure}

\subsubsection{Question 2}
\begin{figure}[H]
\begin{center}
%\framebox[4.0in]{$\;$}
\includegraphics[width=1\linewidth]{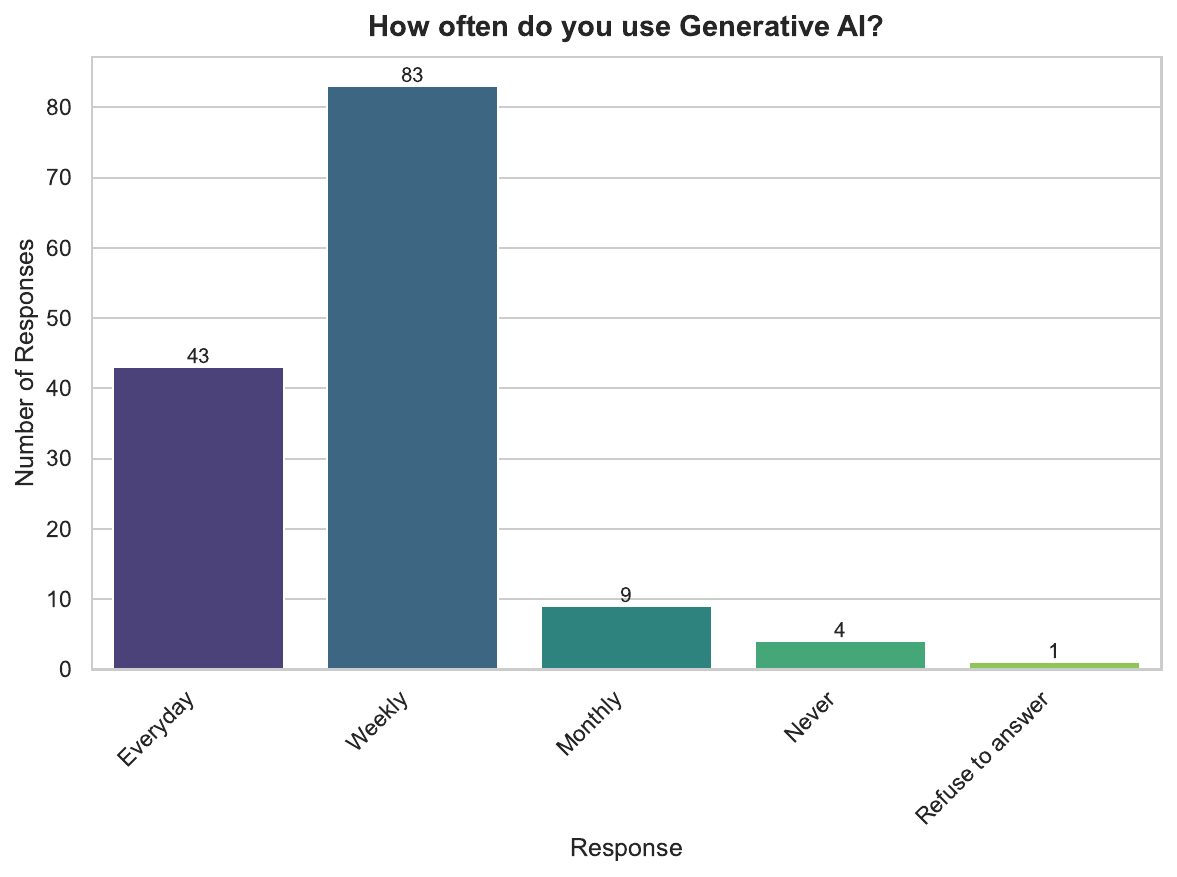}
% \fbox{\rule[-.5cm]{0cm}{4cm} \rule[-.5cm]{4cm}{0cm}}
\end{center}
\caption{Frequency in use of AI (e.g. ChatGPT / Midjourney)}
\label{app:surveyq2}
\end{figure}

\subsubsection{Question 3}
\begin{figure}[H]
\begin{center}
%\framebox[4.0in]{$\;$}
\includegraphics[width=1\linewidth]{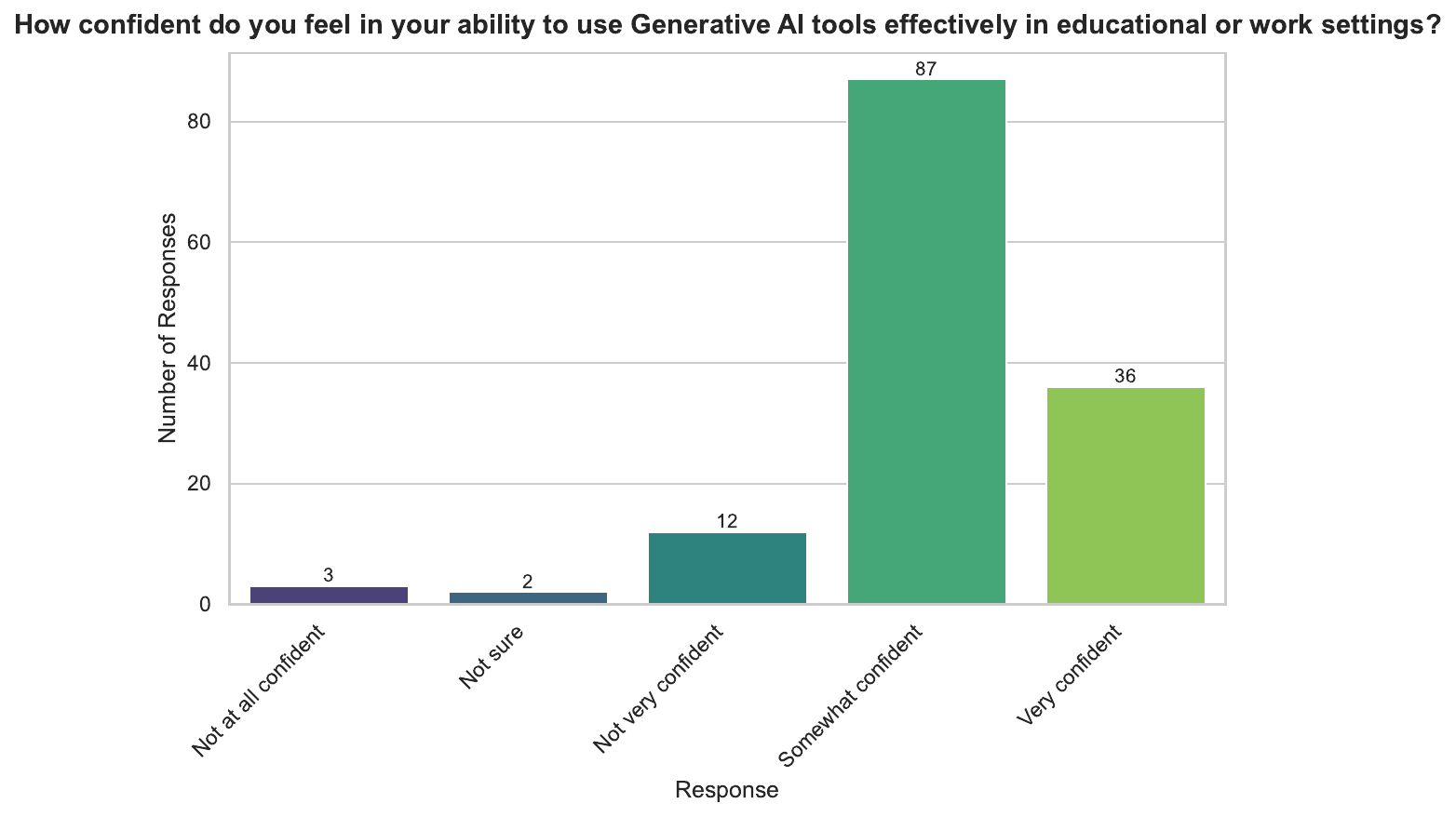}
% \fbox{\rule[-.5cm]{0cm}{4cm} \rule[-.5cm]{4cm}{0cm}}
\end{center}
\caption{Confidence in use of AI in work/ educational settings}
\label{app:surveyq3}
\end{figure}

\subsubsection{Question 4}
\begin{figure}[H]
\begin{center}
%\framebox[4.0in]{$\;$}
\includegraphics[width=0.9\linewidth]{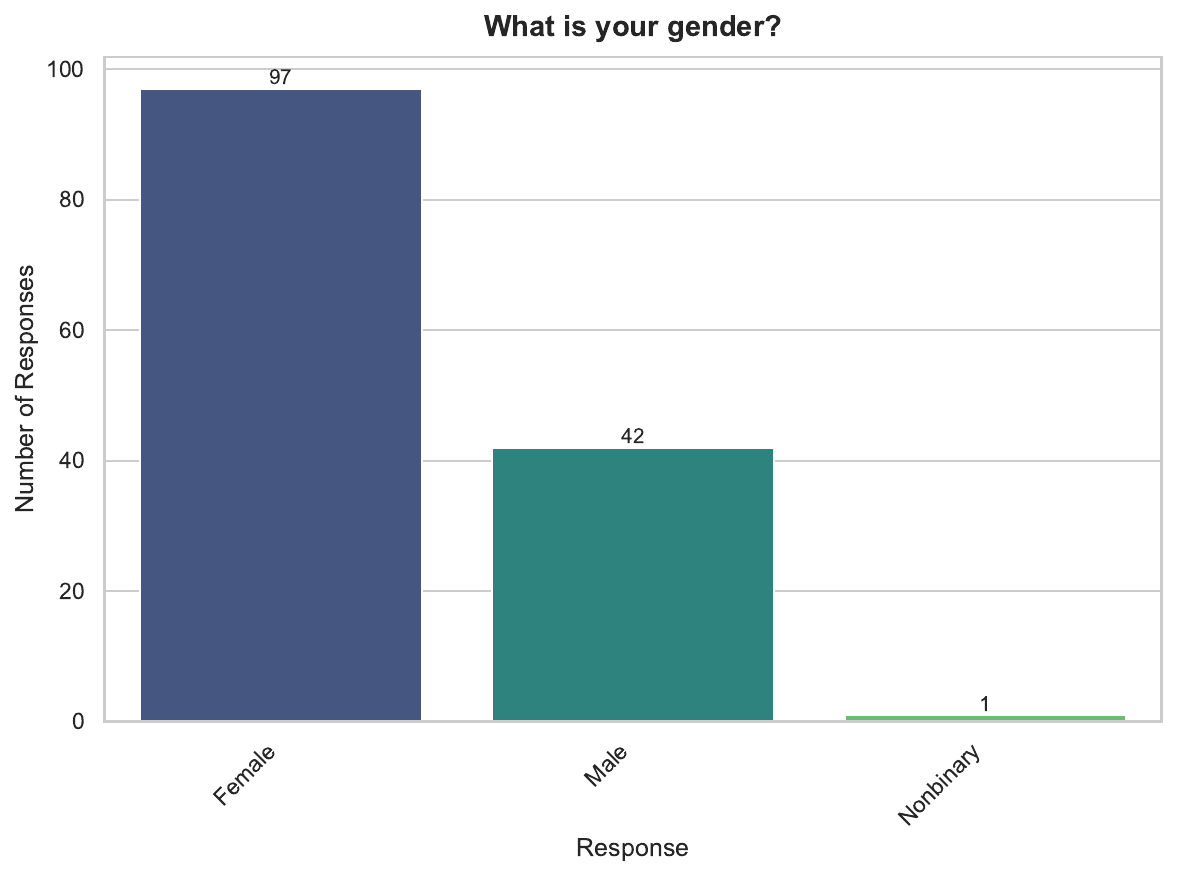}
% \fbox{\rule[-.5cm]{0cm}{4cm} \rule[-.5cm]{4cm}{0cm}}
\end{center}
\label{app:surveyq4}
\caption{Gender}
\end{figure}

\end{document}